\documentclass{aip-cp}

\usepackage[numbers]{natbib}
\usepackage{rotating}
\usepackage{graphicx}
\usepackage{color}
\usepackage[dvipsnames]{xcolor} 

\begin{document}

\title{Neutron Star Merger Remnants: Braking Indices, Gravitational Waves, and the Equation Of State}

\author[aff1,aff2]{Paul D. Lasky\corref{cor1}}
\author[aff1,aff2]{Nikhil Sarin}
\author[aff1,aff2]{Greg Ashton}

\affil[aff1]{Monash Centre for Astrophysics, School of Physics and Astronomy, Monash University, VIC 3800, Australia}
\affil[aff2]{OzGrav: The ARC Centre of Excellence for Gravitational-wave Discovery, Monash University, VIC 3800, Australia}
\corresp[cor1]{Corresponding author: paul.lasky@monash.edu}

\maketitle

\begin{abstract}
The binary neutron star merger GW170817/GRB170817A confirmed that at least some neutron star mergers are the progenitors of short gamma-ray bursts.  Many short gamma-ray bursts have long-term x-ray afterglows that have been interpreted in terms of post-merger millisecond magnetars---rapidly rotating, highly magnetised, massive neutron stars.  We review our current understanding of millisecond magnetars born in short gamma-ray bursts, focusing particularly three main topics.  First, whether millisecond magnetars really do provide the most plausible explain for the x-ray plateau.  Second, determining and observing the gravitational-wave emission from these remnants.  Third, determining the equation of state of nuclear matter from current and future x-ray and gravitational-wave measurements.
\end{abstract}

\section{INTRODUCTION}
Observations of many short- and long-gamma ray bursts show evidence for ongoing energy injection lasting $\sim10^3$--$10^7$ s following the initial burst.  The leading explanation for these long-lived x-ray plateaus is that of a millisecond magnetar; a rapidly-rotating, highly magnetised neutron star~\cite{Dai1998,Zhang2001}.  Further interest in this has again been sparked by the recent observation of the binary neutron star merger GW170817/GRB 170817A with both gravitational waves and across the electromagnetic spectrum~\cite{Abbott2017_GW170817Discovery,Abbott2017_GW170817EM}.  Although there was no definitive evidence for central energy injection in this case, it did confirm that binary neutron star mergers are the progenitors of at least some short gamma-ray bursts.  Moreover, it also renewed hope that a future coincident gravitational-wave and electromagnetic observation will yield further insight into one of the most violent events in the Universe through the discovery of gravitational waves from the post-merger remnant.  No such post-merger gravitational waves were observed following GW170817~\cite{Abbott2017_GW170817PostMerger1,Abbott2017_GW170817PostMerger2,Abbott2017_GW170817PostMerger3}.

Observing and understanding long- or short-lived neutron stars produced from the merger of two neutron stars has the potential to provide invaluable insight into the equation of state of matter at supranuclear densities.  One of the most fundamental predictions from a given equation of state is the maximum mass of non-rotating neutron stars.  There are four possible outcomes from a neutron star merger, which depend primarily on the mass of the progenitors and the equation of state.  Therefore, while insight into the equation of state can be gained from short gamma-ray burst afterglows, as we discuss below, even more can be gained when one has a handle on the progenitor masses by measuring their inspiral with gravitational waves.  Unfortunately, the state of the remnant is highly uncertain~\cite[e.g. see][and references therein]{Margalit2017,Ai2018,Yu2018, Piro2019}

The four potential outcomes for a binary neutron star merger are the prompt collapse to form a black hole, a short-lived hypermassive neutron star that collapses to form a black hole in $\lesssim1$ s, a longer-lived supramassive star that collapses to a black hole on a secular timescale, or an infinitely stable neutron star remnant.  Here, we only consider the last two of these.  

A supramassive neutron star is one that is more massive than the maximum, non-rotating neutron star mass, but is supported from collapse through centrifugal support due to the rapid rotation.  As the star spins down through electromagnetic braking or gravitational-wave emission, it loses angular momentum and therefore eventually loses the centrifugal support to hold it back from collapsing and forming a black hole.  On the other hand, if the neutron star has mass less than that of its maximum, non-rotating mass, it will be infinitely stable and continue to spin down.  Observationally, while the neutron star is spinning down it is expected to emit predominantly x-rays.  The collapse of the supramassive star will cause the x-rays to shut off quickly, causing a steep decay in the x-ray light curve.  

Examples of the above light curves have been observed numerous times, including both long-lived and collapsing systems~\cite{Rowlinson2010,Rowlinson2013,Lu2015}.  Here, we revisit the analysis of many of these systems, focusing particularly on the following three questions.  Firstly, does the data really support the existence of a millisecond magnetar? Second, what gravitational-wave emission can one expect from such long-lived remnants, and do we have any hope of seeing these with second- or third-generation gravitational-wave detectors?  Third, what can we say about the nuclear equation of state from current x-ray and potential gravitational-wave observations.

\subsection{THE GENERALISED MILLISECOND MAGNETAR MODEL}
The original millisecond magnetar model~\cite{Dai1998,Zhang2001} proposed that the star spun down through vacuum dipole radiation.  In pulsar terminology, the braking index $n$ was fixed to a value of three in the torque equation that describes the spindown of the nascent neutron star
\begin{equation}
	\frac{d\Omega}{dt}=-k\Omega^n,\label{eqn:torque}
\end{equation}
where $\Omega$ is the star's angular frequency, and $k$ is a constant of proportionality.  This assumption leads to a prediction that the late-time behaviour of the luminosity goes as $L\propto t^{-2}$.  Equation (\ref{eqn:torque}) should equally apply to isolated radio pulsars, however not one of the well-measured braking indices of such isolated pulsars is consistent with $n=3$; all but one have $n\lesssim3$~\cite[e.g,][]{Archibald2016,Clark2016,Marshall2016}.  

A generalisation of the magnetar model to include an arbitrary braking index has been developed and fit to two x-ray light curves from short gamma-ray bursts~\cite{Lasky2017,Sarin2019}---a point we discuss in considerable detail below.  This model has since been applied to the new fast x-ray transient CDF-S XT2, believed to be a protomagnetar born from a neutron star merger~\cite{Xue2019,Xiao2019}, and long gamma-ray bursts~\cite{Stratta2018,Lu2019a,Mus2019}, and even been generalised to include evolving braking indices that models the evolution of the magnetic field inclination angle and x-ray efficiency~\cite{Mus2019}.  

The rotational kinetic energy of a rigidly rotating body is $E=I\Omega^2/2$, where $I$ is the star's moment of inertia.  As a star spins down, it loses this rotational kinetic energy; the time derivative of this equation therefore gives the rate of change of energy loss.  If one supposes a constant fraction $\eta$ of this energy is converted into x rays, the x-ray luminosity can be expressed as $L=-\eta\Omega (d\Omega/dt)$.  Integrating and using Eqn.~(\ref{eqn:torque}) implies
\begin{equation}
	L=L_0\left(1+\frac{t}{\tau}\right)^{\frac{1+n}{1-n}},\label{eqn:luminosity}
\end{equation}
where $L_0\equiv\eta Ik\Omega_0^{1+n}$ is the initial luminosity, $\Omega_0$ is the initial angular frequency, and $\tau=\Omega_0^{1-n}/\left[(n-1)k\right]$ is the spin-down timescale of the system.  

Equation~(\ref{eqn:luminosity}) includes the characteristic plateau $L=L_0$ for $t\ll\tau$ familiar from the standard magnetar model and a power-law decay $L\propto t^{(1+n)/(1-n)}$ for $t\gg\tau$.  Moreover, when $n=3$ one recovers the usual electromagnetic spin down timescale $\tau=3c^3I/(B_p^2R^6\Omega_0^2)$, where $B_p$ is the poloidal magnetic field, and $R$ the stellar radius~\cite{Lasky2017}.  

\subsection{Case studies: GRBs 130603B and 140903A}
\begin{figure}[h]
  \centerline{\includegraphics[width=350pt]{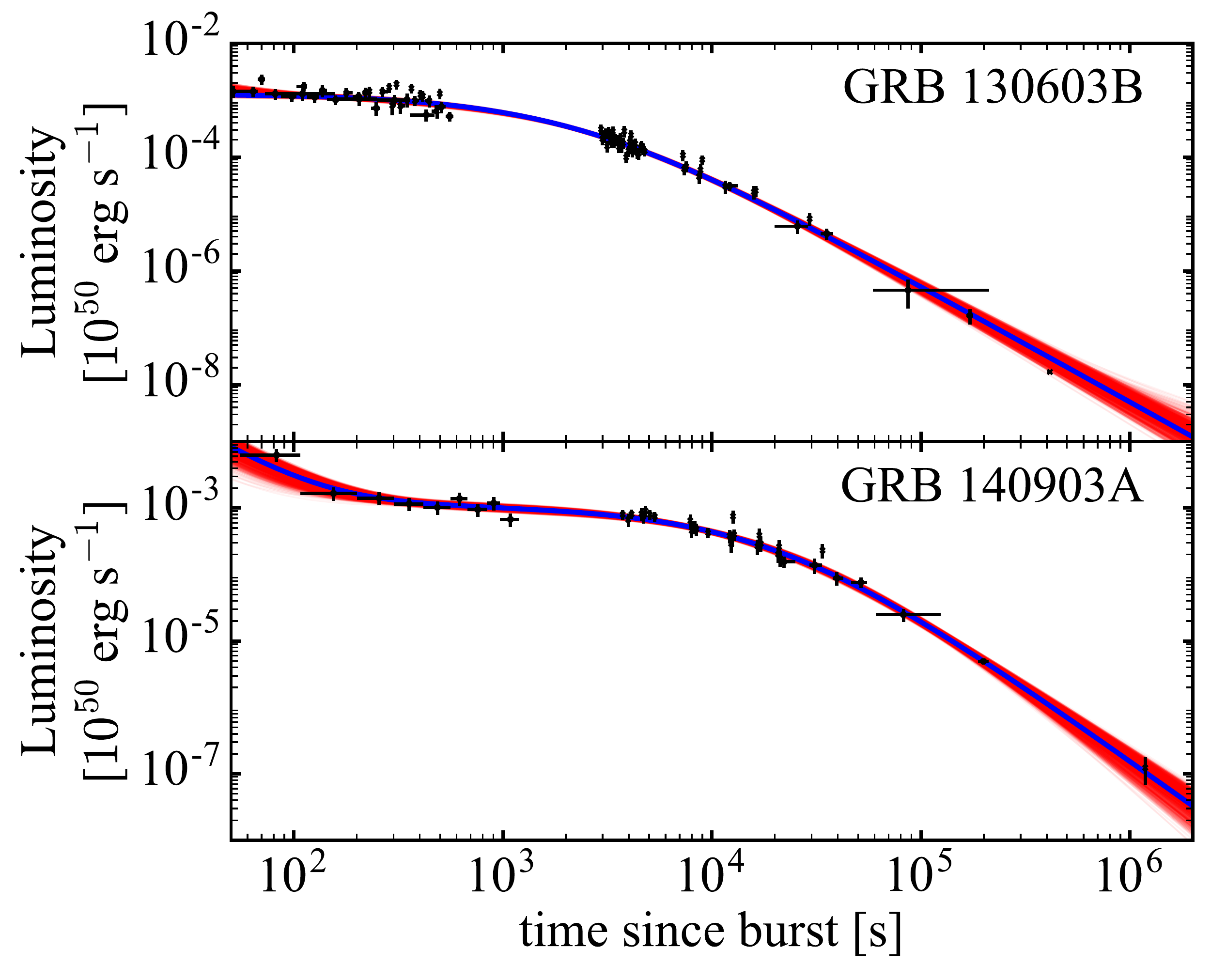}}\label{fig:lightcurves}
  \caption{X-ray light curves for gamma-ray bursts GRB 130603B (top panel) and 140903A (bottom panel).  The black data points show the x-ray data from \textit{Swift's XRT}, \textit{XMM-Newton}, and \textit{Chandra}.  The solid blue curves are the best-fit, generalised millisecond magnetar model including the contribution from an initial power-law decay.  The dark red bands are the superposition of many light curve models drawn from the posterior samples.  Figure reproduced from \citet{Lasky2017}.}
\end{figure}

In the following, we concentrate on two short gamma-ray burst examples; GRB130603B and GRB140903A.  These two gamma-ray bursts are selected as they have well-sampled (and well-behaved) x-ray light curves from the Neil Gehrels \textit{Swift} telescope~\cite[XRT;][]{Burrows2005}, as well as late-time data points from \textit{XMM-Newton} and the \textit{Chandra x-ray observatory}, respectively.  Figure~\ref{fig:lightcurves} shows the two light curves as the black data points.  The late-time data points act as an anchor for the measurement of the braking index; gamma-ray bursts without these late-time points will have significantly less-well-constrained braking indices.  We will present a full catalogue of braking indices measurements in subsequent work~\cite{Sarin2019b}.  

\subsubsection{GRB130603B}
The gamma-ray burst GRB130603B was the first credible detection of a kilonova~\cite{Berger2013, Tanvir2013}.  The standard, $n=3$ millisecond magnetar model has been used to fit both XRT and {\textit XMM} data~\cite{Fan2013,deUgartePostigo2014,Fong2014}, and the more general millisecond magnetar model allowing for a variable braking index; Eqn.~(\ref{eqn:luminosity})~\cite{Lasky2017,Sarin2019}.  The top panel of Fig.~\ref{fig:lightcurves} shows the light curve of GRB 130603B as the black data points, with the maximum posterior model shown as the blue curve.  Note that we fit the model described by Eqn.~(\ref{eqn:luminosity}), together with an initial power law that fits the prompt emission using Bayesian inference.  The dark red band in Fig.~\ref{fig:lightcurves} shows the superposition of many light curves, where each is drawn from a single posterior sample, indicating the uncertainty of the model given the data.  

\begin{figure}[h]
  \centerline{\includegraphics[width=350pt]{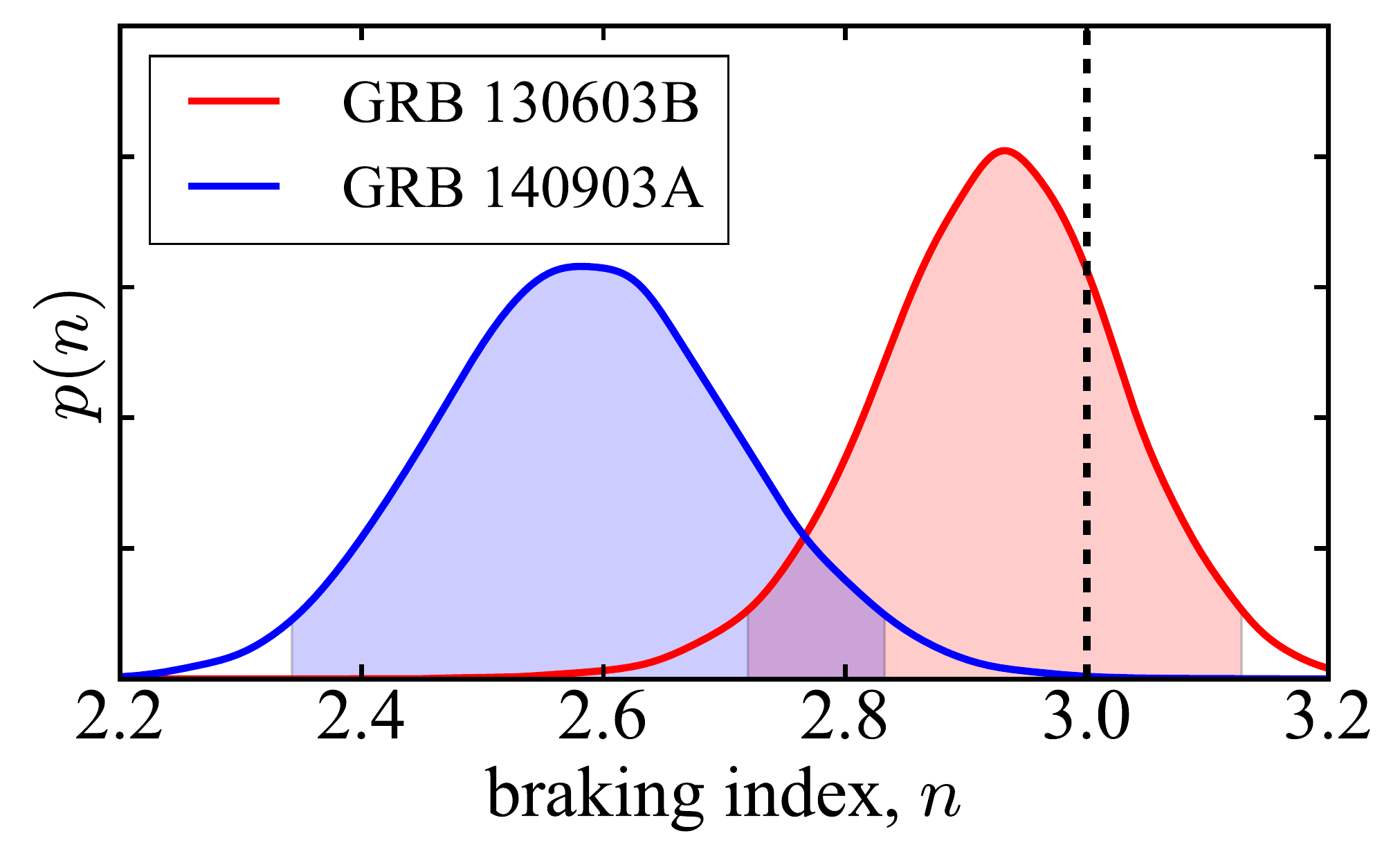}}\label{fig:braking}
  \caption{The braking indices of millisecond magnetars born in gamma-ray bursts GRB 130603B (red) and GRB 140903A (blue).  We show the one-dimensional marginalised posterior distributions, with the shaded regions representing the two-sigma confidence intervals.  The black vertical line is the fiducial $n=3$ value.  Figure reproduced from~\citet{Lasky2017}.}
\end{figure}

In Fig.~\ref{fig:braking} we show the marginalised posterior distribution for the braking index determined by the model in red.  The shaded region shows the two-sigma confidence interval, and the dashed black line shows the fiducial value of $n=3$.  The braking index for this GRB is measured to be $n=2.9\pm0.1$ at one-sigma confidence, which is consistent with the fiducial value.  For a plot of the full posterior including all parameters in the model, see Ref.~\cite{Lasky2017}.

\subsubsection{GRB140903A}
The data for GRB140903A~\cite{Troja2016} has been fit using a fireball model in the absence of additional energy injection from a magnetar~\cite{Troja2016}, as well as with a normal and generalised magnetar model~\cite{Zhang2017,Lasky2017,Sarin2018}.  We note here that these original papers~\cite{Troja2016,Zhang2017} do not discuss the relative goodnesses-of-fit for each model, and therefore do not make quantitative statements about the likelihood of the existence or lack thereof of a millisecond magnetar born in the gamma-ray burst.  This was the motivation for the work introduced in~\citet{Sarin2019}; a point we return to in detail in the following section.

The bottom panel of Fig.~\ref{fig:lightcurves} shows the x-ray light curve of GRB 140903A (black points), together with the maximum posterior model fit (blue curve) of Eqn.~(\ref{eqn:luminosity}) with an initial power-law decay for the prompt emission.  The dark red band again shows the superposition of many light curves, each drawn from a single posterior sample.  One can again see that the late-time data points at $\gtrsim10^5$ and $10^6$ s after the initial burst act to constrain significantly the braking index.  In Fig.~\ref{fig:braking} we show the one-dimensional marginalised posterior distribution for the braking index in blue, which numerically is $n=2.6\pm0.1$ (one-sigma confidence); the fiducial value of $n=3$ is ruled out with 99.95\% confidence.  This last result is not entirely surprising given no radio pulsar with a well-measured braking index is consistent with $n=3$, either, and realistic models of pulsar magnetospheres suggest theoretically that braking indices due to electromagnetic-dominated spin down should be less than three~\cite[e.g.,][]{Melatos1997}.  For a discussion of the physical significance of this result, see Ref.~\cite{Lasky2017}.

\subsection{Can we be sure they are millisecond magnetars?}\label{sec:ModelSelection}
In independent works, the data for GRB 140903A was fit both with and without the energy injection from a central engine~\cite{Troja2016,Zhang2017,Lasky2017,Sarin2018}.  These seemingly contradictory results motivate a robust model selection study, which was first published in~\citet{Sarin2019}.  

Bayesian model selection can be used to calculate the relative preference data has for any two models.  In the present case, we want to test between a model with energy injection from a magnetar central engine and without; the generalised millisecond magnetar model of~Eqn.(\ref{eqn:luminosity}) and the fireball model.  The latter is typically written as a combination of $N$ power laws, and model selection can again be used to determine the optimal number of power laws that fit to a given light curve.  In reality, there are a number of extra subtelties associated with this calculation, but we defer the reader to Ref.~\cite{Sarin2019} for details; here we simply discuss the most salient features of the model-selection process.

The Bayes factor is simply the ratio of Bayesian evidences ${\rm BF}_{1/2}\equiv\mathcal{Z}_1/\mathcal{Z}_2$ for each of two models.  For both GRB 130603B and GRB 140903A, we found that the Bayes factor favours four power-law components for the fireball model~\cite{Sarin2019}.  Overwhelmingly, model selection between this four-component fireball model and the generalised millisecond magnetar model prefers the latter for both gamma-ray burst afterglow light curves.  In particular, the Bayes factor for magnetar versus best-fit fireball ${\rm BF}_{\rm M/F}=19$ and 2271 for GRB 130603B and GRB 140903A, respectively~\cite{Sarin2019}.  Assuming each hypothesis is equally likely \textit{a priori}, this tells us that the data prefers the magnetar model by 19 and 2271 times over the fireball model, respectively.

But this is not the full story.  Do we really think the two hypotheses are \textit{a priori} equally likely?  Physically, we always expect the presence of a fireball, but we do not always expect the existence of a magnetar-driven outflow, simply because not all short gamma-ray bursts are expected to produce remnant magnetars.  Statistically, this implies we need to introduce a prior odds $\Pi_{\rm M}/\Pi_{\rm F}$, which quantifies our relative expectation for the magnetar and fireball scenarios.  Then, it is not the Bayes factor that quantifies which model is preferred by the data, but it is instead the \textit{odds} $\mathcal{O}$, given by
\begin{equation}
    \mathcal{O}_{M/F}=\frac{\Pi_{\rm M}}{\Pi_{\rm F}}\frac{\mathcal{Z}_{\rm M}}{\mathcal{Z}_{\rm F}}.\label{eqn:odds}
\end{equation}

We can calculate the prior odds, and hence derive the odds, as follows.  Let's assume all short gamma-ray bursts progenitors are binary neutron star mergers.  The existence of a long-lived post-merger millisecond magnetar is then dependent mainly on the progenitor masses and the equation of state of nuclear matter.  The latter sets an upper limit on the mass of non-rotating neutron stars, a quantity known as the Tolman-Oppenheimer-Volkoff (TOV) mass, which is empirically unknown up to certain lower limits.  Indeed, observations of galactic neutron stars show the maximum mass exceeds $M_{\rm TOV}\ge2.01\pm0.04$~\cite{Antoniadis2013}, although see the interesting new measurement of a pulsar with mass $M=2.17^{+0.11}_{-0.10}$ at one-sigma confidence~\cite{Cromartie2019}.

For a rapidly-rotating neutron star to be stable for more than~$\sim1$ s, its mass must be below~$\lesssim1.2\times\,M_{\rm TOV}$~\cite{Cook1994}.  When the mass of the star is above $M_{\rm TOV}$, the star can be temporarily held up from collapse by the extra support centrifugal forces provide.  Such a \textit{supramassive} star born from neutron-star mergers will collapse~$\lesssim5\times10^4$~s following the merger~\cite{Ravi2014}.  The x-ray light curves of GRBs 130603B and 140903A show no signs of rapid decay indicating collapse to a black hole did not happen for more than a few times $10^5$~s following the initial gamma-ray burst, implying the remnant magnetar, if formed, was not in the supramassive regime but was infinitely stable.  Therefore, if a magnetar was born in either of these gamma-ray bursts, their mass was less than $M_{\rm TOV}$.  

With this information in hand, we only need to know the mass of the neutron star progenitors to estimate the fraction of binary neutron star mergers that result in infinitely stable neutron star remnants.  We can estimate this probabilistically following the calculation of~\citet{Lasky2014}.  In particular, we assume the mass distribution of galactic double neutron star systems is representative of the universal population, and use the mass distribution $M=1.32\pm0.11\,M_\odot$ outlined in~\citet{Kiziltan2013} (although see the more recent works of~\cite{Keitel2019,Farrow2019}).  To find the post-merger mass we demand conservation of rest mass, and invoke two limits for the mass ejected during the merger.  The first provides an upper limit on the mass distribution of the remnant by assuming no mass is ejected throughout the merger, which gives the post-merger mass distribution as $p(M)=2.45\pm0.14\,M_\odot$.  But we can also consider a more aggresive scenario in which the quantity of mass ejected in the merger as inferred from the kilonova observations of GW170817 are representative of typical binary neutron star mergers.  This yields a post-merger mass distribution of $p(M)=2.38\pm0.14\,M_\odot$.  For details of these calculations, see~\citet{Sarin2019}.

\begin{figure}[h]
  \centerline{\includegraphics[width=350pt]{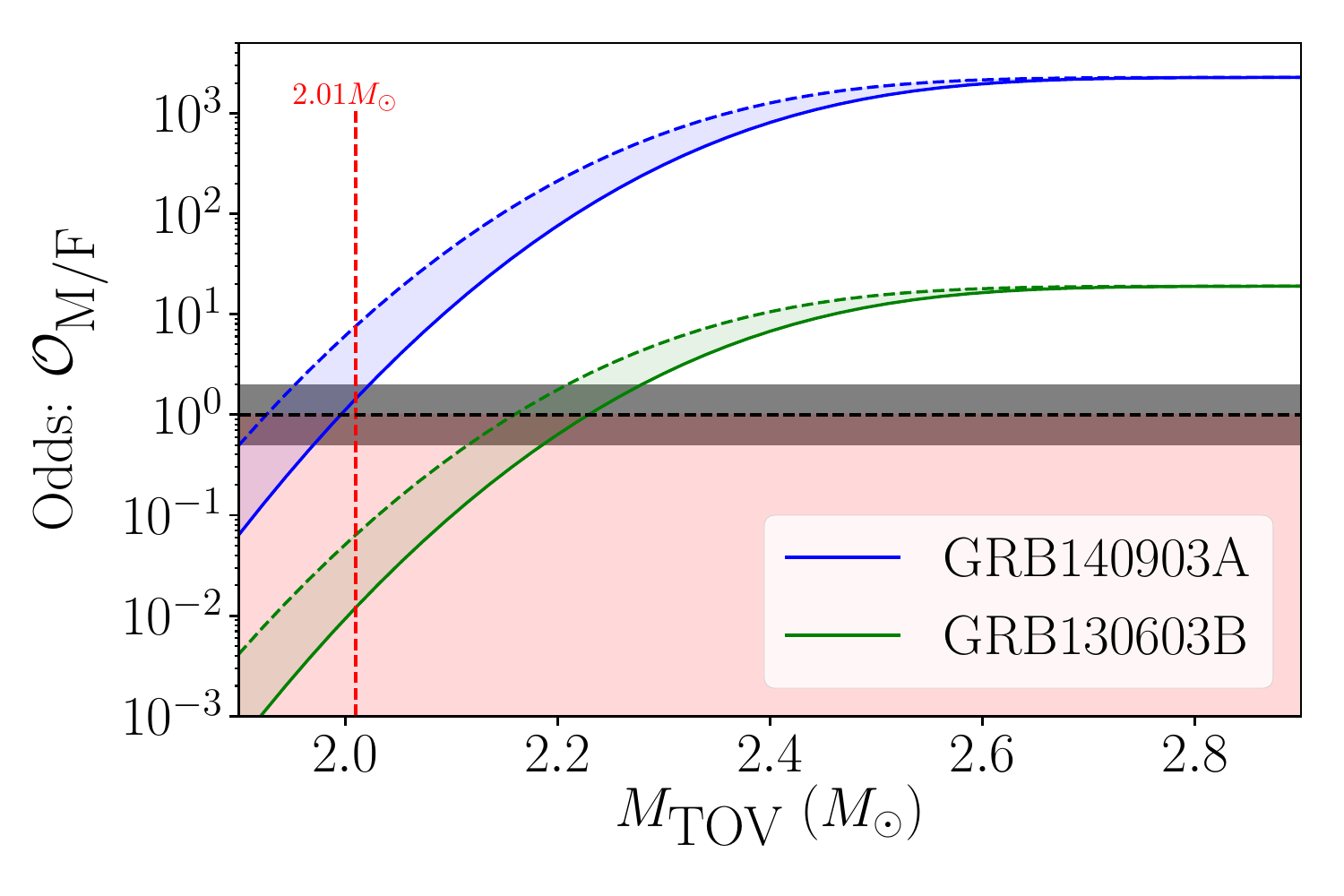}}
  \caption{Odds for magnetar versus fireball model $\mathcal{O}_{\rm M/F}$ as a function of the maximum, non-rotating neutron star mass $M_{\rm TOV}$ for two short gamma-ray bursts from \citet{Sarin2019}.  The solid curves ignore mass ejected through the merger, while the dashed curves assume 0.07 $M_\odot$ is ejected, consistent with the kilonova observations of GW170817 \cite[e.g.,][]{Evans2017}.  The red shaded regions is where the fireball model is favoured by the data over the magnetar model.}\label{fig:odds}
\end{figure}

The prior odds for a magnetar versus fireball is simply the probability that the mass of the post-merger remnant is less than $M_{\rm TOV}$:
\begin{equation}
    \frac{\Pi_{\rm M}}{\Pi_{\rm F}}=\int_0^{M_{\rm TOV}}p\left(M\right)dM.\label{eqn:oddsratio}
\end{equation}
Equations~(\ref{eqn:odds}) and~(\ref{eqn:oddsratio}) can then be used to calculate the odds as a function of the unknown $M_{\rm TOV}$, which is shown in Fig.~\ref{fig:odds}.  This figure shows the odds increasing as a function of $M_{\rm TOV}$; for larger values of the maximum neutron star mass, one is more likely to form a stable neutron star remnant from a binary collision.  

Figure~\ref{fig:odds} shows that, for all viable values of the TOV mass $M_{\rm TOV}\gtrsim2.01\,M_\odot$, and even assuming the most conservative scenario of zero mass loss during merger, the data for GRB 140903A prefers the magnetar model over the fireball model.  We can definitively say that, in this case, a magnetar was born in the gamma-ray burst.  

We do not have the same level of certainty for GRB 130603B; if the TOV mass is $M_{\rm TOV}\gtrsim2.3\,M_\odot$, then the odds $\mathcal{O}_{\rm M/F}\gtrsim2$, implying the magnetar model is prefered by the data.  However, the converse is also true; if the TOV mass is smaller than $M_{\rm TOV}\lesssim2.2\,M_\odot$, the data prefers the fireball model, implying a magnetar central engine is not required to explain the x-ray light curve.  In the TOV mass range between $\sim2.2\,M_\odot$ and $2.3\,M_\odot$, the data is not informative towards either model.

\section{GRAVITATIONAL WAVES FROM NEUTRON STAR REMNANTS}
Figure~\ref{fig:odds} emphasises that, at least for the case of one short gamma-ray burst, a millisecond magnetar was born and rapidly spun down for more than $10^6$ s.  Indeed, many more short gamma-ray bursts have been modelled similarly~\cite[e.g.,][]{Rowlinson2013, Lu2015}, although rigorous model selection \`a la Ref.~\cite{Sarin2019} is still forthcoming~\cite{Sarin2019b}.

Given such rapidly rotating, massive magnetars likely exist, what can we say about their gravitational-wave emission?  For GRBs 130603B and 140903A we have determined the braking index $n$ through the torque equation; see Fig.~\ref{fig:braking}.  In general, while a braking index $n=3$ indicates vacuum dipole spin down, a braking index of $n=5$ is indicative of gravitational-wave dominated spin down.  That the two GRBs analysed above and the~\citet{Rowlinson2013} and~\citet{Lu2015} samples of short gamma-ray bursts are consistent with $n\ll5$ implies that, at late times ($t\gtrsim\tau$), the spin down is dominated by electromagnetic rather than gravitational-wave emission~\cite{Lasky2017}.  This allows us to put constraints on the amplitude of gravitational-wave emission from these purported magnetars, and therefore determine the detectability with second- and third-generation gravitational-wave detectors.  

In~\citet{Lasky2016}, we analyzed eight gamma-ray bursts with known redshifts that show good fits to the magnetar model with $n=3$ from~\citet{Rowlinson2013}.  These light curves allow us to place strict upper limits on the gravitational-wave emission; if the gravitational-wave amplitude was sufficiently large, the $n=3$ magnetar model would \textit{not} provide a good fit to the data.  

Fits to the light curve with the $n=3$ magnetar model allow one to estimate the value of the initial spin period of the nascent neutron star as well as it's magnetic field strength~\cite{Rowlinson2013}.  However, these can only be known up to a factor, which is the efficiency $\eta$ of turning spin-down energy into x-ray luminosity~\cite[e.g.,][]{Rowlinson2014}, which is also likely to be time dependent~\cite[e.g.,][]{Mus2019}.  

\begin{figure}[h]
  \centerline{
  \includegraphics[width=220pt]{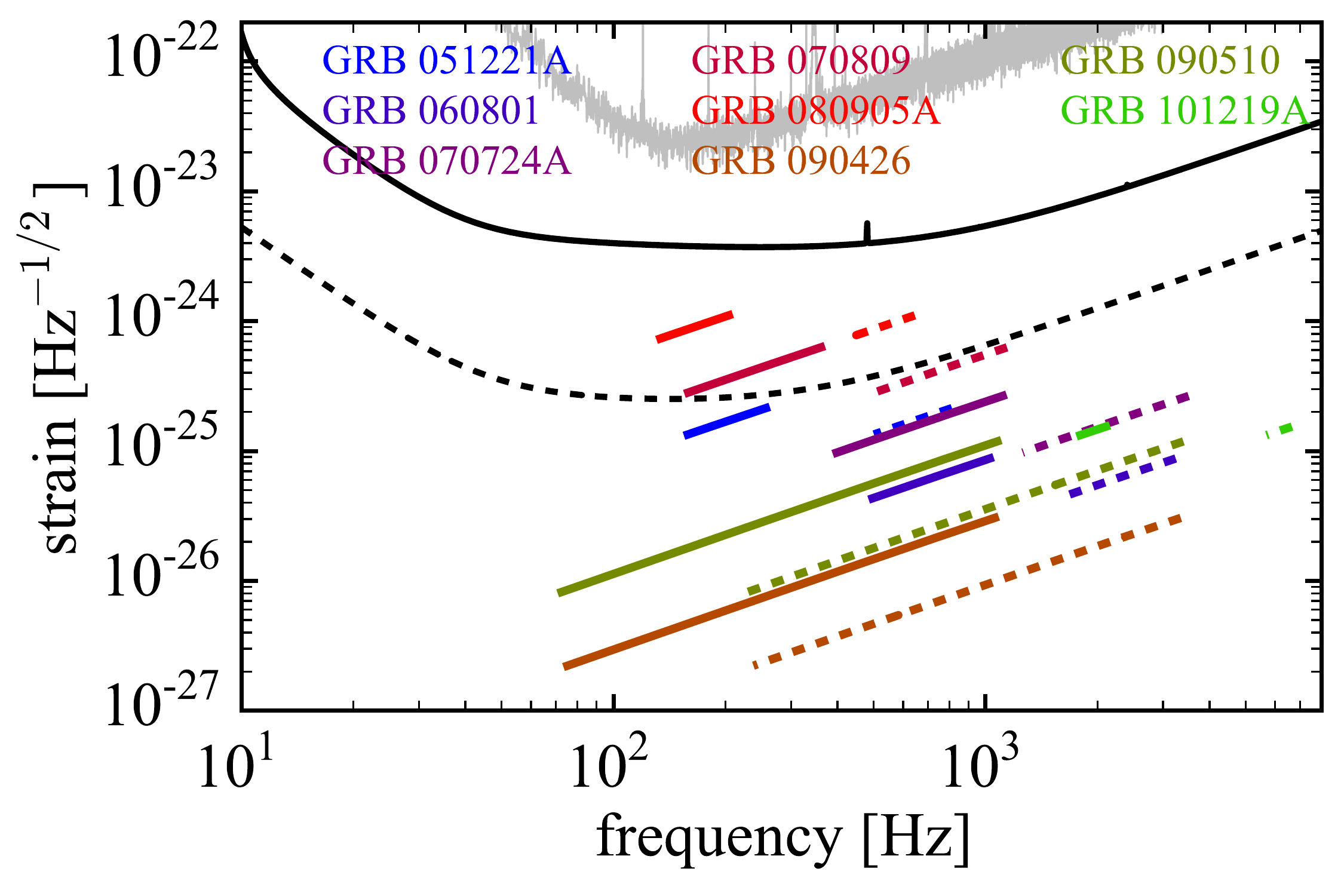}
  \includegraphics[width=220pt]{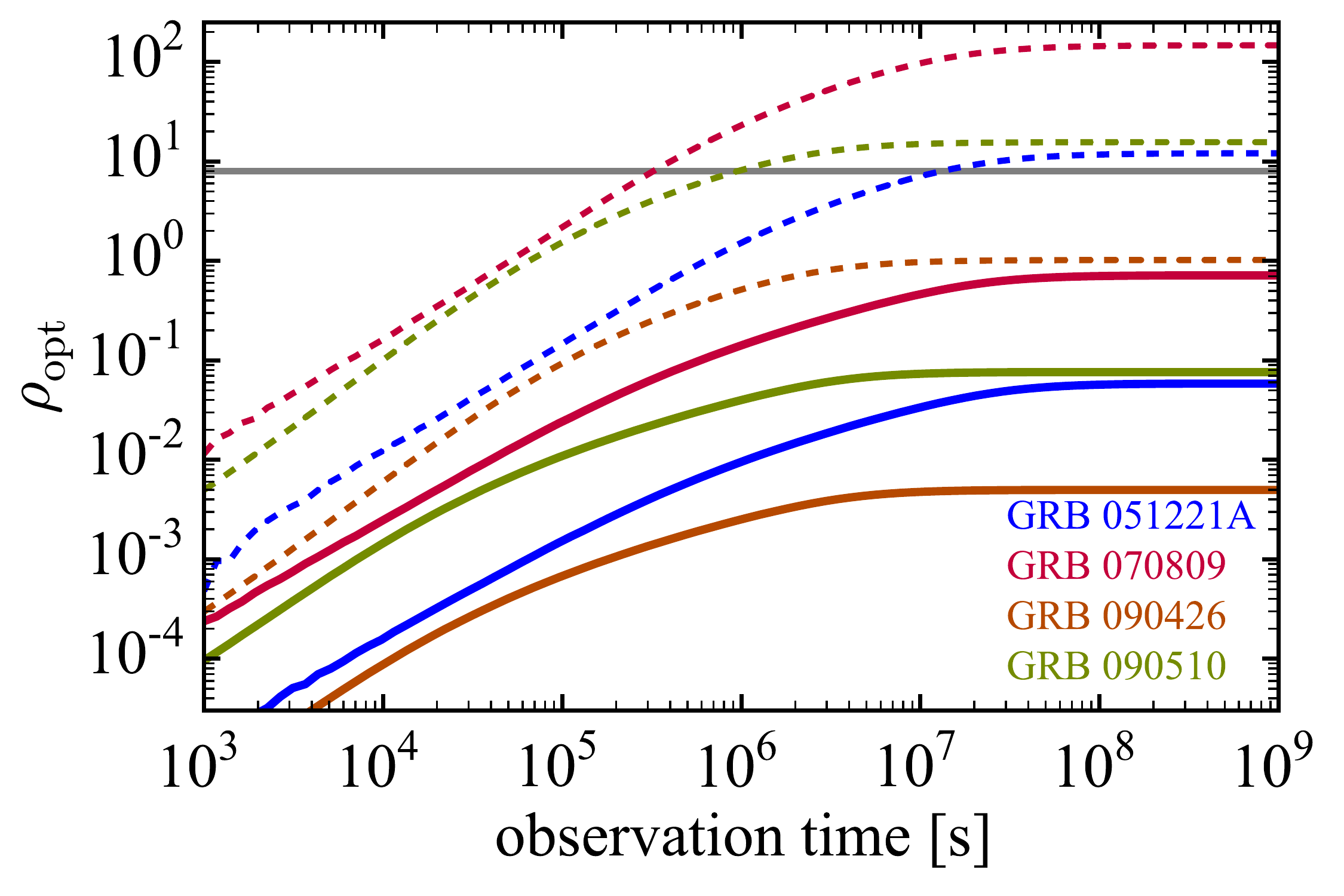}
  }
  \caption{Optimistic predicted upper limits on the gravitational-wave strain (left panel) and detectability (right panel) for eight and four short gamma-ray bursts, respectively.  Figure taken from~\citet{Lasky2016}  Left panel: the gravitational-wave strain is calculated assuming an x-ray efficiency of $\eta=1$ (solid, coloured curves) and $\eta=0.1$ (dashed, coloured curves)---see text for detail---and compared to initial S5 LIGO sensitivity curve (solid grey curve),the projected sensitivity of Advanced LIGO at design sensitivity (solid black curve) and the Einstein Telescope (dashed black curve).  Right panel: the optimal matched filter signal-to-noise ratio $\rho_{\rm opt}$ is calculated for Advanced LIGO at design sensitivity (solid, coloured curves) and the Einstein Telescope (dashed coloured curves).}\label{fig:GWs}
\end{figure}

We take the inferred values of the neutron star initial spin periods and magnetic fields at face value~\cite{Rowlinson2013}, and use these to calculate the maximum gravitational-wave signal.  The left-hand panel of Figure~\ref{fig:GWs} shows these gravitational-wave strain upper limits for the eight gamma-ray bursts assuming $\eta=1$ (solid coloured curves) and $\eta=0.1$ (dashed coloured curves).  We compare these predicted upper limits with strain sensitivity curves from the S5 run of Initial LIGO~\cite[light grey curve;][]{PSD:iLIGO}, Advanced LIGO at design sensitivity~\cite[solid black curve;][]{PSD:ALIGO} and the Einstein Telescope~\cite[dashed black curve;][]{Punturo2010,Hild2011}.  

Four of the gamma-ray bursts in the left-hand panel of Fig.~\ref{fig:GWs} have light curves that show rapid decay at some point in their evolution, indicating the nascent star was born supramassive, and collapsed to form a black hole after many hundreds of seconds~\cite{Rowlinson2013,Ravi2014}---we return to this below.  The other four gamma-ray bursts do not show such behaviour; for these magnetars we calculate the optimal signal-to-noise ratio $\rho_{\rm opt}$ as a function of observation time for the strain curves shown in the left-hand panel of Fig.~\ref{fig:GWs}.  The right-hand panel shows $\rho_{\rm opt}$ as a function of observation time for Advanced LIGO at design sensitivity~\cite[solid coloured curves;][]{PSD:ALIGO} and the Einstein Telescope~\cite[dashed coloured curve;][]{Hild2011}.

The curves in the right-hand panel of Fig.~\ref{fig:GWs} are interesting for two reasons.  First, they show a characteristic power-law increase for some time before plateauing.  This plateau is due to the fact that the neutron star has sufficiently spun down by that time that the gravitational-wave emission is minimal.  It implies that gravitational-wave observations of such systems reach the point of minimal returns after some nominal timescale; we return to this in the next section.  Second, we note that the signal-to-noise ratio for these systems is rather pessimistic, and even with the most optimistic scenario and unrealistic detection pipelines (see discussion in Ref.~\cite{Lasky2016}), these systems could only be marginally detectable with third-generation telescopes such as the Einstein Telescope.

But the authors of Ref.~\cite{Lasky2016} were ignorant at the time of writing.  They were under the misguided impression that binary neutron stars did not merge in the local Universe.  In August of 2017, this was shown to be wrong with the discovery of GW170817 at a distance of just $\sim40$ Mpc~\cite{Abbott2017_GW170817Discovery, Abbott2017_GW170817EM}, implying detections of gravitational waves from binary neutron star post-merger remnants become significantly more likely.  As the strain scales inversely with the distance, and the signal-to-noise ratio is proportional to distance $h\propto1/d\propto\rho_{\rm opt}$, one can simply scale the results presented in Fig.~\ref{fig:GWs}, and realise that, under the most optimistic assumptions presented in~\citet{Lasky2016}, a detection of gravitational waves from a long-lived post-merger remnant at $\sim40$ Mpc is possible for Advanced LIGO operating at design sensitivity.

The results presented in~\citet{Lasky2016} and scaled to nearby sources are derived by simply looking at late-time ($\gtrsim100$ s) x-ray light curves following gamma-ray bursts.  For this reason, they are an upper limit, but they say nothing of the possible mechanisms that could even cause such gravitational-wave emission.  Numerous authors have attempted to answer the question of how large a gravitational-wave amplitude one would expect from a long-lived post-merger remnant.  Loosely, such emission mechanisms include magnetic-field induced stellar ellipticities~\cite{Bonazzola1996, Cutler2002}, bar modes~\cite{Lai1995}, and $r$ modes~\cite{Andersson1998}, but the estimates for the gravitational-wave amplitude vary wildly~\cite[e.g.,][]{Corsi2009,Fan2013,DallOsso2015,Doneva2015,Lasky2016,Gao2017,Dai2019}.

But there is one further piece of evidence that limits the total gravitational-wave emission from post-merger remnants, if our understanding of the x-ray plateaus is correct.  Consider magnetic-field induced ellipticities as one of the example gravitational-wave mechanisms above, although the argument holds for the other mechanisms as well.  Naturally, one can detect sources at greater distance if their ellipticity is higher, implying larger amplitude of gravitational waves.  However, in practice, one can rule out arbitrarily high ellipticities as the total energy emitted in gravitational waves would exceed that available in the from the total rotational energy budget of the system~\cite{Sarin2018}.

\begin{figure}[h]
  \centerline{
  \includegraphics[width=220pt]{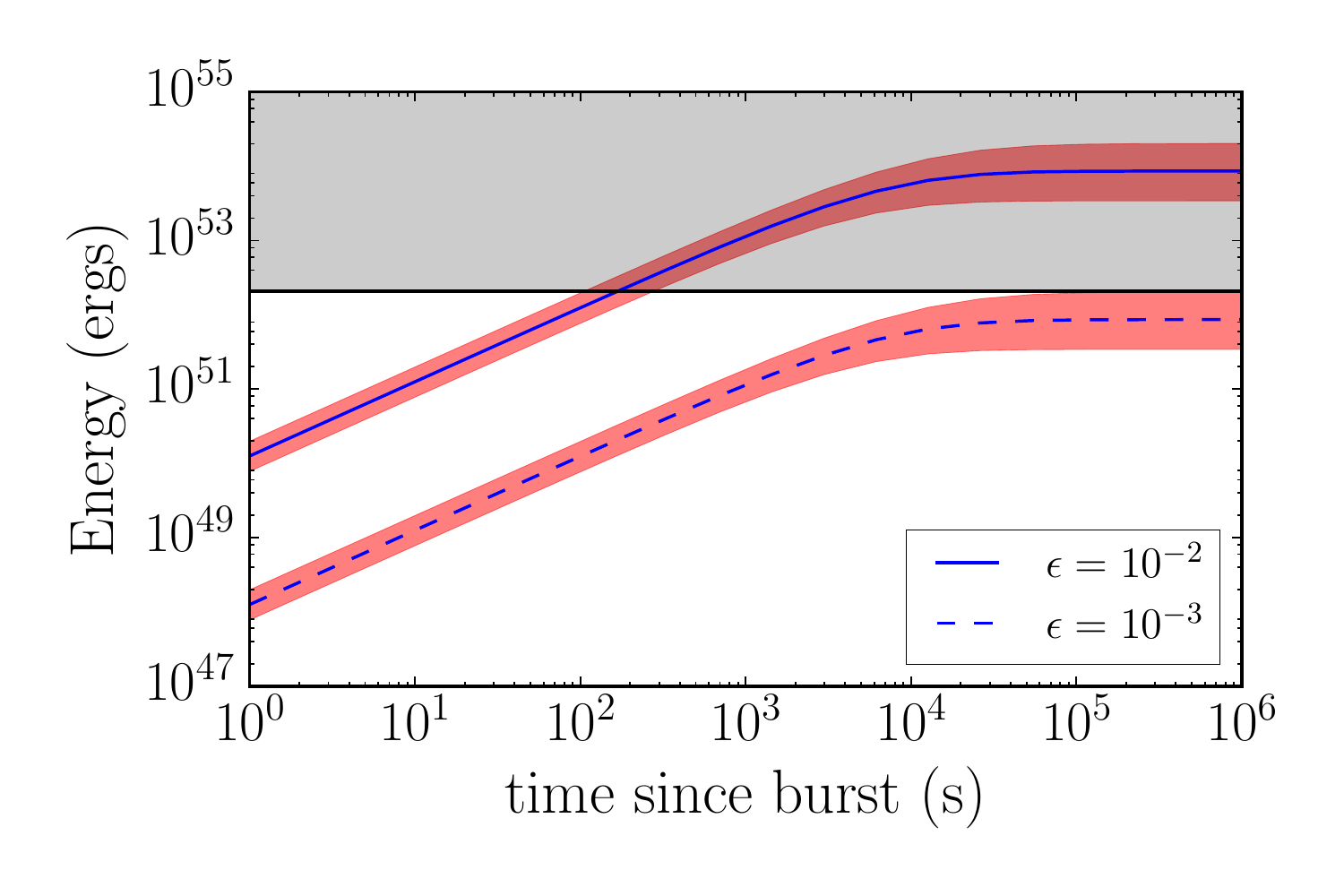}
  \includegraphics[width=220pt]{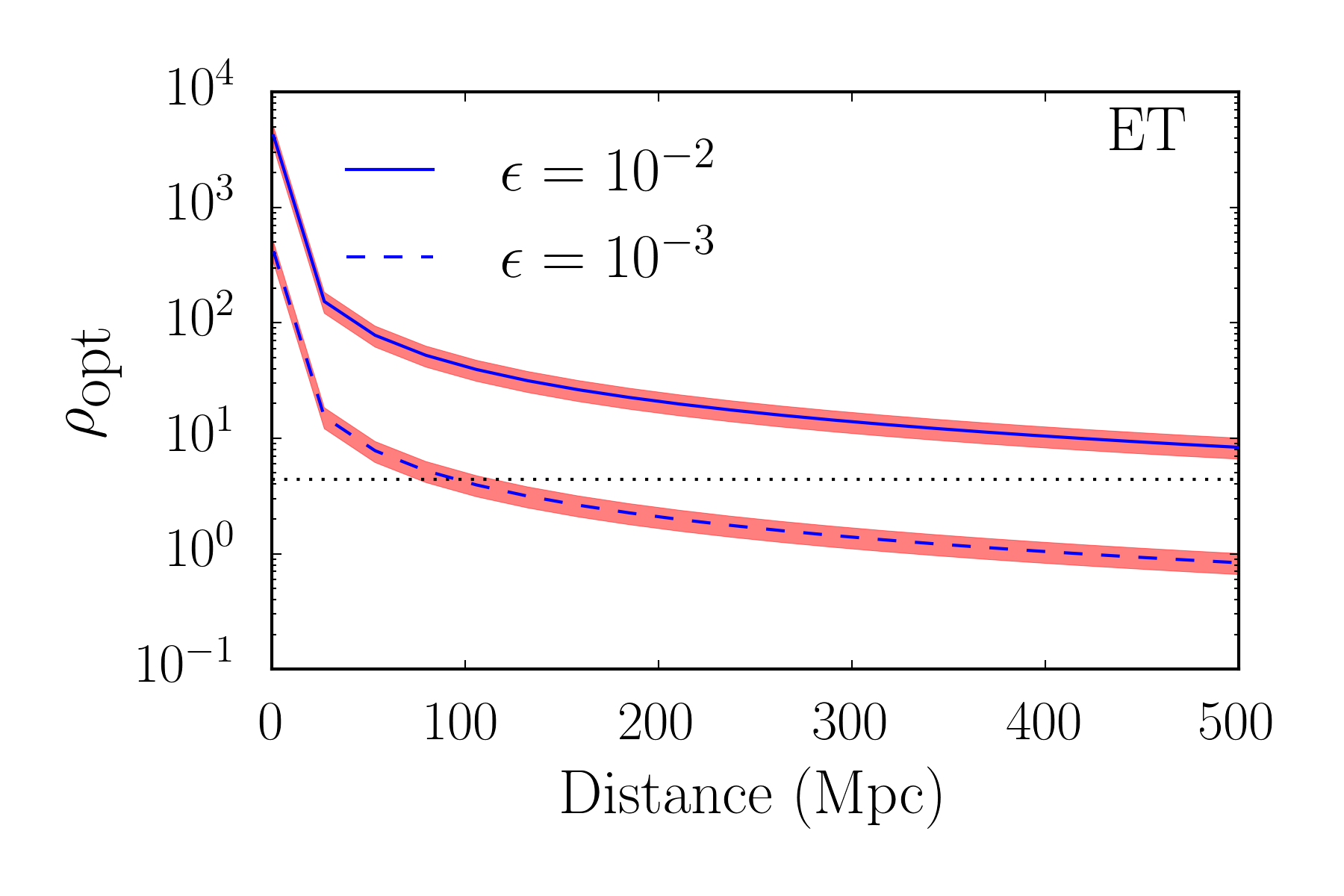}
  }
  \caption{Left: Gravitational-wave energy emitted by a millisecond magnetar with GRB 140903A-like properties, and two different values of ellipticity.  The solid black line represents the rotational energy budget of the system; if the blue/red curves go into the shaded region, the parameter space is unphysical.  Right: Optimal matched filter signal-to-noise ratio for the same system as a function of distance assuming an Einstein Telescope sensitivity curve.  The horizontal dashed line represents a detection threshold.  A GRB 140903A-like system with ellipticity $\epsilon=10^{-3}$ can therefore only be seen out to $\sim45$ Mpc.  This plot assumes an optimistic observation time of $5\times10^4$ s.  Both plots are taken from~\citet{Sarin2018}.}\label{fig:EnergyBudget}
\end{figure}

To illustrate this point, we fit the generalised magnetar model (Eqns.~\ref{eqn:torque} and~\ref{eqn:luminosity}) to the light curve from the gamma-ray burst GRB 140903A and derive a model for gravitational-wave emission based on the parameters inferred from the fit; for details, see~\citet{Sarin2018}.  We then calculate the total gravitational-wave energy emitted as a function of time since the burst, and plot this for two example values of ellipticity in the left-hand plot of Fig.~\ref{fig:EnergyBudget}.  In that figure, we also plot the total rotational energy budget available to the system assuming the derived initial spin period of the system as the horizontal black line.  The shaded region above that line represents an unphysical region of parameter space; if the blue/red curves go into the shaded region, then more gravitational-wave energy has been emitted than was available in rotational energy.  This figure highlights that arbitrarily large ellipticities are not possible and, for the case of GRB 140903A, one cannot have an ellipticity above $\epsilon\gtrsim10^{-3}$~\cite{Sarin2018}. 

We will put this derived upper limit on the ellipticity into a physical context in a subsequent section.  In terms of gravitational-wave detectability, one can convert such an ellipticity, using the derived magnetar-model parameters, into an optimal matched-filter signal-to-noise ratio $\rho_{\rm opt}$, and ask the question: at what distance could I hope to detect such a signal.  In the right-hand panel of Fig.~\ref{fig:EnergyBudget}, we plot $\rho_{\rm opt}$ for the Einstein Telescope as a function of distance, assuming a relatively optimistic observing time of $5\times10^4$ s (see the discussion in Ref.~\cite{Sarin2018} as to why this is optimistic).  This plot shows such a gravitational-wave signal could be seen by the Einstein Telescope out to a distance of approximately 45 Mpc --- note that the $\epsilon=10^{-2}$ systems can be seen out to much larger distances, however they reside in an unphysical region of parameter space as explained previously.  A similar plot to the right-hand panel of Fig.~\ref{fig:EnergyBudget} can be seen assuming Advanced LIGO operating at design sensitivity in~\citet{Sarin2018}; the unfortunate punchline there is that such systems can only be seen out to~$\sim2$ Mpc.

While a somewhat pessimistic picture has so far been painted, we do point out one potential optimistic avenue for detecting gravitational-wave emission from long-lived merger remnants.  The light-curve analyses presented to this point rely on information predominantly gained at times $t\gtrsim\tau$, where $\tau$ is the spin-down timescale of the system, and corresponds to the knee of the light curves seen in Fig.~\ref{fig:lightcurves}.  For GRBs 130603B and 140903A, $\tau=(2.2\pm0.1)\times10^3$ s and $(2.0\pm0.2)\times10^4$ s, respectively~\cite{Lasky2017}.  We certainly not able to say anything about the evolution of these systems for the first~$\sim100$ s as \textit{Swift} generally takes that long to slew into position.  Moreover, even if the systems are initially gravitational-wave spin-down dominated for $t\lesssim\tau$, this would hardly show up as a deviation in the light curve as a gravitational-wave dominated x-ray afterglow still has a plateau phase in which the luminosity is largely constant~\cite{Zhang2001,Lasky2016}.

While the above suggests we do not have direct information about the early evolution of these neutron stars, there may be tentative evidence that their spin down is, in fact, dominated by gravitational-wave emission at early times, and only becomes electromagnetically-dominated at late times.  The evidence comes from looking at those millisecond magnetars born in short gamma-ray bursts that are seen to collapse via a sharp decay feature in their x-ray plateau.  \citet{Ravi2014} predicted the spin down evolution of those nascent neutron stars in a bid to understand what their collapse-time distribution \textit{should} look like on theoretical grounds --- see also~\cite{Piro2017}.  Indeed, if one assumes these stars spin down only through magnetic vacuum dipole radiation, then the observed distribution of collapse times does not match the predicted distribution~\cite{Ravi2014}.  Many authors have speculated on the nature of this disagreement, with all outcomes being equally intriguing.  For example, this could be evidence that the post-merger remnant is actually a quark star~\cite{Li2016,Li2017,Drago2018}, or in fact gravitational-wave energy is taking angular momentum away from the system, causing them to collapse earlier~\cite{Gao2016, Lu2017,Lin2019}.  

With this in mind, the most likely known binary neutron star merger to have gravitational-wave emission detectable by the current generation of gravitational-wave instruments is GW170817/GRB170817A.  And so, the only remaining question is: were gravitational waves from the post-merger remnant of GW170817 detected?

\subsection{Were gravitational waves from the post-merger remnant of GW170817 detected?}
No~\cite{Abbott2017_GW170817PostMerger1,Abbott2017_GW170817PostMerger2,Abbott2017_GW170817PostMerger3}.

\section{DETERMINING THE EQUATION OF STATE FROM NEUTRON STAR REMNANTS}
We finally turn to attempts to measure the equation of state of nuclear matter from the x-ray plateau observations of short gamma-ray bursts.  The idea is simple; some subset of short gamma-ray burst afterglows exhibit steep decays, which has been interpreted as the collapse of the supramassive magnetar to form a black hole~\cite{Rowlinson2013}.  Fitting the original, $n=3$ millisecond magnetar model to these light curves prior to their collapse time $t_{\rm col}$, one can estimate the birth spin period of the magnetar $p_0$ as well as the magnetic field strength $B_p$.  If the magnetar has a mass $M_{\rm TOV} <m\lesssim1.2M_{\rm TOV}$, then it is considered to be supramassive, and will collapse to form a black hole when it has spun down sufficiently that it can no longer support it's own mass through the extra support centrifugal forces supply.  In that case, one can derive an equation for the collapse time~\cite{Lasky2014}
\begin{equation}
    t_{\rm col}=\frac{3c^3I}{4\pi^2B_p^2R^6}\left[\left(\frac{m-M_{\rm TOV}}{\alpha M_{\rm TOV}}\right)^{2/\beta}-p_0^2\right]. \label{eqn:tcol}
\end{equation}
Here, $R$ and $I$ are the radius and moment of inertia of the neutron star.  The parameters $\alpha$ and $\beta$ are equation-of-state dependent parameters that define a neutron stars maximum spin period such that~\cite{Lyford2003}
\begin{equation}
    m_{\rm max}(p)\approx M_{\rm TOV}\left(1+\alpha p^\beta\right).\label{eqn:mmax}
\end{equation}
\citet{Lasky2014} calculated $\alpha$ and $\beta$ for a series of five representative equations of state by calculating equilibrium sequences of $m_{\rm max}$; those values can be found in Ref.~\cite{Ravi2014}.

Equation~(\ref{eqn:tcol}) depends on a number of parameters which can be divided into three categories: those that can be indirectly observed using the x-ray light curves, $t_{\rm col}$, $p_0$, and $B_{p}$; those that relate to the equation of state of nuclear matter, $M_{\rm TOV}$, $\alpha$, $\beta$, $R$ and $I$; and the mass of the remnant, $m$.  It may seem strange that the radius and moment of inertia are put into the category of equation of state parameters, but if we know the mass of the star, then the equation of state tells us what the radius and moment of inertia are.  The task for determining the equation of state given an observation of a collapsing supramassive neutron star observed via an x-ray afterglow comes down to determining the actual mass of the neutron star remnant.  If that is known, then Eqns.~(\ref{eqn:tcol}) and~(\ref{eqn:mmax}) can be solved to get the equation of state parameters.

\begin{figure}[h]
  \centerline{
  \includegraphics[width=220pt]{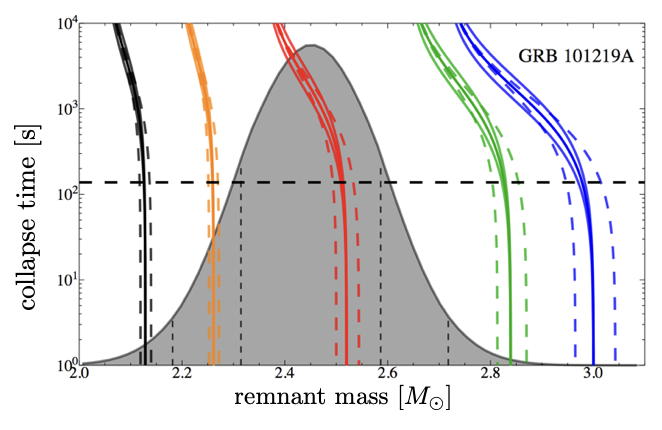}
  \includegraphics[width=220pt]{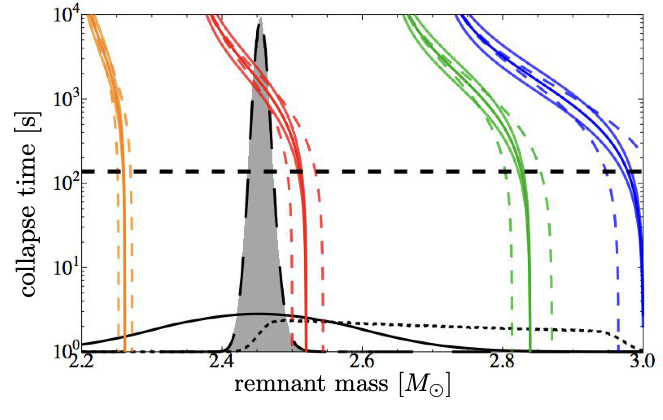}
  }
  \caption{Collapse time predictions for short gamma-ray burst GRB 101219A.  The coloured curves correspond to predicted collapse times given a remnant mass for five different equations of state (see the text for the details about the equation of state), and the horizontal dashed line corresponds to the observed collapse time. The mass distribution in the left panel is that derived from just looking at the mass distribution of known galactic binary neutron star systems.  The mass distribution on the right-hand side also combines information from a hypothetical, future gravitational-wave observation.  Figures taken from~\citet{Lasky2014}.}\label{fig:EOS}
\end{figure}

There are two potential ways to get the mass of a given progenitor.  The first is to look at the known, galactic population of binary neutron star systems, which have a relatively tight mass distribution.  Taking that mass distribution~\cite{Kiziltan2013}, and demanding conservation of rest mass, one can derive a distribution for the remnant mass---see~\citet{Lasky2014} for details---which is plotted as the grey distribution on the left-hand panel of Fig.~\ref{fig:EOS} (the height of the distribution is arbitrary, and does not correspond to the left-hand axis label).

Over-plotted on the left-hand panel are five different solutions of Eqn.~(\ref{eqn:tcol}) for parameters derived for short gamma-ray burst GRB 101219A assuming different remnant masses $m$ for five representative equations of state.  Explicitly, these are SLy (black curves; $M_{\rm TOV}=2.05\,M_\odot$), APR (orange curves; $M_{\rm TOV}=2.20\,M_\odot$), GM1 (red curves; $M_{\rm TOV}=2.37\,M_\odot$), AB-N (green curves; $M_{\rm TOV}=2.67\,M_\odot$), and AB-L (blue curves; $M_{\rm TOV}=2.71\,M_\odot$).  The horizontal dashed curve is the observed collapse time from Ref.~\cite{Rowlinson2013}.  

The left-hand panel of Fig.~\ref{fig:EOS} is to be interpreted as follows.  Consider for example the AB-L equation of state (blue curve); if this is the true equation of state, then to explain this collapse time, the remnant mass had to be $m\approx3.0\,M_\odot$, which is a significant outlier from the posterior mass distribution plotted in grey.  Stated another way, if this equation of state were true, then the overwhelmingly most-likely scenario is that a stable, rather than supramassive, neutron star would have been born from this collision.  From this plot alone, one can all but rule out equations of state with such high TOV masses.  It is worth emphasising that these equation-of-state curves assume the spin down of the star is dominated by electromagnetic torques rather than gravitational-wave emission; as eluded to earlier, this may not be true.

Figure~\ref{fig:EOS} shows only one representative example of a single gamma-ray burst.  There are four gamma-ray bursts presented in~\citet{Lasky2014}, and many more in~\citet{Lu2015}.  The take-away message is that we can rule out exotic equations of state with very large TOV masses $M_{\rm TOV}\gtrsim2.7\,M_\odot$.  Perhaps more interestingly, to explain all of the short gamma-ray burst observations, we also need the TOV mass to be above a certain minimum threshold of $M_{\rm TOV}\gtrsim2.2$.  

The right-hand panel shows another way in which we can determine the mass of the remnant, but it relies on detecting the inspiral through gravitational waves and coincidentally seeing a long-lived x-ray plateau with steep decay.  If we see one of these in the future, then the grey distribution shows a hypothetical mass posterior for the remnant mass.  This is derived by combining the mass distribution from the galactic binary neutron stars described previously (solid black distribution in the right-hand panel) with the hypothetical remnant mass distribution inferred from the gravitational-wave observation (dotted black curve).  The resultant mass distribution is significantly less broad, and therefore significantly enhances our ability to determine the equation of state---see Ref.~\cite{Lasky2014} for details.

\section{CONCLUSIONS}
Binary neutron star mergers and their subsequent evolution has the potential to provide valuable insight into fundamental physics and astrophysics.  The emission of both high-energy photons and gravitational waves from the post-merger remnant are being studied in detail, both from a theoretical and observational perspective.  While our current understanding from electromagnetic observations is impressive, a multi-messenger gravitational-wave plus electromagnetic observation of a post-merger remnant has the potential to be truly revolutionary.  With Advanced LIGO and Virgo reaching even better sensitivity in the third LIGO observing run, KAGRA joining soon, and the prospect for further future upgrades and LIGO-India joining the network, the future of this field is exciting.

\section{ACKNOWLEDGMENTS}
We thank James Clark for useful comments on the manuscript.  PDL is extremely grateful to the organisers of the conference for the invitation and for hosting a wonderful meeting.  PDL is supported through Australian Research Council Future Fellowship FT160100112 and ARC Discovery Project DP180103155.  NS is supported through an Australian Postgraduate Award.

\bibliographystyle{aipnum-cp}%
\bibliography{LaskyXiamenProceedings.bib}%

\begin{thebibliography}{67}%
\makeatletter
\providecommand \@ifxundefined [1]{%
 \@ifx{#1\undefined}
}%
\providecommand \@ifnum [1]{%
 \ifnum #1\expandafter \@firstoftwo
 \else \expandafter \@secondoftwo
 \fi
}%
\providecommand \@ifx [1]{%
 \ifx #1\expandafter \@firstoftwo
 \else \expandafter \@secondoftwo
 \fi
}%
\providecommand \natexlab [1]{#1}%
\providecommand \enquote  [1]{``#1''}%
\providecommand \bibnamefont  [1]{#1}%
\providecommand \bibfnamefont [1]{#1}%
\providecommand \citenamefont [1]{#1}%
\providecommand \href@noop [0]{\@secondoftwo}%
\providecommand \href [0]{\begingroup \@sanitize@url \@href}%
\providecommand \@href[1]{\@@startlink{#1}\@@href}%
\providecommand \@@href[1]{\endgroup#1\@@endlink}%
\providecommand \@sanitize@url [0]{\catcode `\$12\catcode `\&12\catcode
  `\#12\catcode `\^12\catcode `\_12\catcode `\%12\relax}%
\providecommand \@@startlink[1]{}%
\providecommand \@@endlink[0]{}%
\providecommand \url  [0]{\begingroup\@sanitize@url \@url }%
\providecommand \@url [1]{\endgroup\@href {#1}{\urlprefix }}%
\providecommand \urlprefix  [0]{URL }%
\providecommand \Eprint [0]{\href }%
\providecommand \doibase [0]{http://dx.doi.org/}%
\providecommand \selectlanguage [0]{\@gobble}%
\providecommand \bibinfo  [0]{\@secondoftwo}%
\providecommand \bibfield  [0]{\@secondoftwo}%
\providecommand \translation [1]{[#1]}%
\providecommand \BibitemOpen [0]{}%
\providecommand \bibitemStop [0]{}%
\providecommand \bibitemNoStop [0]{.\EOS\space}%
\providecommand \EOS [0]{\spacefactor3000\relax}%
\providecommand \BibitemShut  [1]{\csname bibitem#1\endcsname}%
\let\auto@bib@innerbib\@empty
\bibitem [{\citenamefont {Dai}\ and\ \citenamefont {Lu}(1998)}]{Dai1998}%
  \BibitemOpen
  \bibfield  {author} {\bibinfo {author} {\bibfnamefont {Z.~G.}\ \bibnamefont
  {Dai}}\ and\ \bibinfo {author} {\bibfnamefont {T.}~\bibnamefont {Lu}},\
  }\href@noop {} {\bibfield  {journal} {\bibinfo  {journal} {A\&A}\ }\textbf
  {\bibinfo {volume} {333}},\ p.\ \bibinfo {pages} {L87} (\bibinfo {year}
  {1998})}\BibitemShut {NoStop}%
\bibitem [{\citenamefont {Zhang}\ and\ \citenamefont
  {M{\'{e}}sz{\'{a}}ros}(2001)}]{Zhang2001}%
  \BibitemOpen
  \bibfield  {author} {\bibinfo {author} {\bibfnamefont {B.}~\bibnamefont
  {Zhang}}\ and\ \bibinfo {author} {\bibfnamefont {P.}~\bibnamefont
  {M{\'{e}}sz{\'{a}}ros}},\ }\href@noop {} {\bibfield  {journal} {\bibinfo
  {journal} {Astrophys. J. L.}\ }\textbf {\bibinfo {volume} {552}},\
  p.~\bibinfo {pages} {35} (\bibinfo {year} {2001})}\BibitemShut {NoStop}%
\bibitem [{\citenamefont {{Abbott}}\ \emph
  {et~al.}(2017{\natexlab{a}})\citenamefont {{Abbott}} \emph
  {et~al.}}]{Abbott2017_GW170817Discovery}%
  \BibitemOpen
  \bibfield  {author} {\bibinfo {author} {\bibfnamefont {B.~P.}\ \bibnamefont
  {{Abbott}}} \emph {et~al.},\ }\href@noop {} {\bibfield  {journal} {\bibinfo
  {journal} {Phys. Rev. Lett.}\ }\textbf {\bibinfo {volume} {119}},\ p.\
  \bibinfo {pages} {161101} (\bibinfo {year} {2017}{\natexlab{a}})}\BibitemShut
  {NoStop}%
\bibitem [{\citenamefont {{Abbott}}\ \emph
  {et~al.}(2017{\natexlab{b}})\citenamefont {{Abbott}} \emph
  {et~al.}}]{Abbott2017_GW170817EM}%
  \BibitemOpen
  \bibfield  {author} {\bibinfo {author} {\bibfnamefont {B.~P.}\ \bibnamefont
  {{Abbott}}} \emph {et~al.},\ }\href@noop {} {\bibfield  {journal} {\bibinfo
  {journal} {Astrophys. J. L.}\ }\textbf {\bibinfo {volume} {848}},\ p.\
  \bibinfo {pages} {L12} (\bibinfo {year} {2017}{\natexlab{b}})}\BibitemShut
  {NoStop}%
\bibitem [{\citenamefont {{Abbott}}\ \emph
  {et~al.}(2017{\natexlab{c}})\citenamefont {{Abbott}} \emph
  {et~al.}}]{Abbott2017_GW170817PostMerger1}%
  \BibitemOpen
  \bibfield  {author} {\bibinfo {author} {\bibfnamefont {B.~P.}\ \bibnamefont
  {{Abbott}}} \emph {et~al.},\ }\href {\doibase 10.3847/2041-8213/aa9a35}
  {\bibfield  {journal} {\bibinfo  {journal} {Astrophys. J. L.}\ }\textbf
  {\bibinfo {volume} {851}},\ p.\ \bibinfo {pages} {L16} (\bibinfo {year}
  {2017}{\natexlab{c}})}\BibitemShut {NoStop}%
\bibitem [{\citenamefont {{Abbott}}\ \emph {et~al.}(2019)\citenamefont
  {{Abbott}} \emph {et~al.}}]{Abbott2017_GW170817PostMerger2}%
  \BibitemOpen
  \bibfield  {author} {\bibinfo {author} {\bibfnamefont {B.~P.}\ \bibnamefont
  {{Abbott}}} \emph {et~al.},\ }\href@noop {} {\bibfield  {journal} {\bibinfo
  {journal} {Astrophys. J.}\ }\textbf {\bibinfo {volume} {875}},\ p.\ \bibinfo
  {pages} {160} (\bibinfo {year} {2019})}\BibitemShut {NoStop}%
\bibitem [{\citenamefont {Abbott}\ \emph {et~al.}(2019)\citenamefont {Abbott}
  \emph {et~al.}}]{Abbott2017_GW170817PostMerger3}%
  \BibitemOpen
  \bibfield  {author} {\bibinfo {author} {\bibfnamefont {B.~P.}\ \bibnamefont
  {Abbott}} \emph {et~al.},\ }\href@noop {} {\bibfield  {journal} {\bibinfo
  {journal} {Phys. Rev. X}\ }\textbf {\bibinfo {volume} {9}},\ p.\ \bibinfo
  {pages} {011001} (\bibinfo {year} {2019})}\BibitemShut {NoStop}%
\bibitem [{\citenamefont {{Margalit}}\ and\ \citenamefont
  {{Metzger}}(2017)}]{Margalit2017}%
  \BibitemOpen
  \bibfield  {author} {\bibinfo {author} {\bibfnamefont {B.}~\bibnamefont
  {{Margalit}}}\ and\ \bibinfo {author} {\bibfnamefont {B.~D.}\ \bibnamefont
  {{Metzger}}},\ }\href@noop {} {\bibfield  {journal} {\bibinfo  {journal}
  {Astrophys. J L.}\ }\textbf {\bibinfo {volume} {850}},\ p.\ \bibinfo {pages}
  {L19} (\bibinfo {year} {2017})}\BibitemShut {NoStop}%
\bibitem [{\citenamefont {{Ai}}\ \emph {et~al.}(2018)\citenamefont {{Ai}} \emph
  {et~al.}}]{Ai2018}%
  \BibitemOpen
  \bibfield  {author} {\bibinfo {author} {\bibfnamefont {S.}~\bibnamefont
  {{Ai}}} \emph {et~al.},\ }\href@noop {} {\bibfield  {journal} {\bibinfo
  {journal} {Astrophys. J.}\ }\textbf {\bibinfo {volume} {860}},\ p.~\bibinfo
  {pages} {57} (\bibinfo {year} {2018})}\BibitemShut {NoStop}%
\bibitem [{\citenamefont {{Yu}}, \citenamefont {{Liu}},\ and\ \citenamefont
  {{Dai}}(2018)}]{Yu2018}%
  \BibitemOpen
  \bibfield  {author} {\bibinfo {author} {\bibfnamefont {Y.-W.}\ \bibnamefont
  {{Yu}}}, \bibinfo {author} {\bibfnamefont {L.-D.}\ \bibnamefont {{Liu}}}, \
  and\ \bibinfo {author} {\bibfnamefont {Z.-G.}\ \bibnamefont {{Dai}}},\
  }\href@noop {} {\bibfield  {journal} {\bibinfo  {journal} {Astrophys. J.}\
  }\textbf {\bibinfo {volume} {861}},\ p.\ \bibinfo {pages} {114} (\bibinfo
  {year} {2018})}\BibitemShut {NoStop}%
\bibitem [{\citenamefont {{Piro}}\ \emph {et~al.}(2019)\citenamefont {{Piro}}
  \emph {et~al.}}]{Piro2019}%
  \BibitemOpen
  \bibfield  {author} {\bibinfo {author} {\bibfnamefont {L.}~\bibnamefont
  {{Piro}}} \emph {et~al.},\ }\href@noop {} {\bibfield  {journal} {\bibinfo
  {journal} {Mon. Not. R. Astron. Soc.}\ }\textbf {\bibinfo {volume} {483}},\
  \unskip\ \bibinfo {pages} {1912--1921} (\bibinfo {year} {2019})}\BibitemShut
  {NoStop}%
\bibitem [{\citenamefont {{Rowlinson}}\ \emph {et~al.}(2010)\citenamefont
  {{Rowlinson}} \emph {et~al.}}]{Rowlinson2010}%
  \BibitemOpen
  \bibfield  {author} {\bibinfo {author} {\bibfnamefont {A.}~\bibnamefont
  {{Rowlinson}}} \emph {et~al.},\ }\href@noop {} {\bibfield  {journal}
  {\bibinfo  {journal} {Mon. Not. R. Astron. Soc.}\ }\textbf {\bibinfo {volume}
  {409}},\ \unskip\ \bibinfo {pages} {531--540} (\bibinfo {year}
  {2010})}\BibitemShut {NoStop}%
\bibitem [{\citenamefont {Rowlinson}\ \emph {et~al.}(2013)\citenamefont
  {Rowlinson}, \citenamefont {O'brien}, \citenamefont {Metzger}, \citenamefont
  {Tanvir},\ and\ \citenamefont {Levan}}]{Rowlinson2013}%
  \BibitemOpen
  \bibfield  {author} {\bibinfo {author} {\bibfnamefont {A.}~\bibnamefont
  {Rowlinson}}, \bibinfo {author} {\bibfnamefont {P.~T.}\ \bibnamefont
  {O'brien}}, \bibinfo {author} {\bibfnamefont {B.~D.}\ \bibnamefont
  {Metzger}}, \bibinfo {author} {\bibfnamefont {N.~R.}\ \bibnamefont {Tanvir}},
  \ and\ \bibinfo {author} {\bibfnamefont {A.~J.}\ \bibnamefont {Levan}},\
  }\href@noop {} {\bibfield  {journal} {\bibinfo  {journal} {Mon. Not. R.
  Astron. Soc.}\ }\textbf {\bibinfo {volume} {430}},\ p.\ \bibinfo {pages}
  {1061} (\bibinfo {year} {2013})}\BibitemShut {NoStop}%
\bibitem [{\citenamefont {L{\"{u}}}\ \emph {et~al.}(2015)\citenamefont
  {L{\"{u}}}, \citenamefont {Zhang}, \citenamefont {Lei}, \citenamefont {Li},\
  and\ \citenamefont {Lasky}}]{Lu2015}%
  \BibitemOpen
  \bibfield  {author} {\bibinfo {author} {\bibfnamefont {H.-J.}\ \bibnamefont
  {L{\"{u}}}}, \bibinfo {author} {\bibfnamefont {B.}~\bibnamefont {Zhang}},
  \bibinfo {author} {\bibfnamefont {W.-H.}\ \bibnamefont {Lei}}, \bibinfo
  {author} {\bibfnamefont {Y.}~\bibnamefont {Li}}, \ and\ \bibinfo {author}
  {\bibfnamefont {P.~D.}\ \bibnamefont {Lasky}},\ }\href@noop {} {\bibfield
  {journal} {\bibinfo  {journal} {Astrophys. J.}\ }\textbf {\bibinfo {volume}
  {805}},\ p.~\bibinfo {pages} {89} (\bibinfo {year} {2015})}\BibitemShut
  {NoStop}%
\bibitem [{\citenamefont {{Archibald}}\ \emph {et~al.}(2016)\citenamefont
  {{Archibald}} \emph {et~al.}}]{Archibald2016}%
  \BibitemOpen
  \bibfield  {author} {\bibinfo {author} {\bibfnamefont {R.~F.}\ \bibnamefont
  {{Archibald}}} \emph {et~al.},\ }\href@noop {} {\bibfield  {journal}
  {\bibinfo  {journal} {Astrophys. J. L.}\ }\textbf {\bibinfo {volume} {819}},\
  p.~\bibinfo {pages} {16} (\bibinfo {year} {2016})}\BibitemShut {NoStop}%
\bibitem [{\citenamefont {{Clark}}\ \emph {et~al.}(2016)\citenamefont {{Clark}}
  \emph {et~al.}}]{Clark2016}%
  \BibitemOpen
  \bibfield  {author} {\bibinfo {author} {\bibfnamefont {C.~J.}\ \bibnamefont
  {{Clark}}} \emph {et~al.},\ }\href@noop {} {\bibfield  {journal} {\bibinfo
  {journal} {Astrophys. J. L.}\ }\textbf {\bibinfo {volume} {832}},\
  p.~\bibinfo {pages} {15} (\bibinfo {year} {2016})}\BibitemShut {NoStop}%
\bibitem [{\citenamefont {{Marshall}}\ \emph {et~al.}(2016)\citenamefont
  {{Marshall}}, \citenamefont {{Guillemot}}, \citenamefont {{Harding}},
  \citenamefont {{Martin}},\ and\ \citenamefont {{Smith}}}]{Marshall2016}%
  \BibitemOpen
  \bibfield  {author} {\bibinfo {author} {\bibfnamefont {F.~E.}\ \bibnamefont
  {{Marshall}}}, \bibinfo {author} {\bibfnamefont {L.}~\bibnamefont
  {{Guillemot}}}, \bibinfo {author} {\bibfnamefont {A.~K.}\ \bibnamefont
  {{Harding}}}, \bibinfo {author} {\bibfnamefont {P.}~\bibnamefont {{Martin}}},
  \ and\ \bibinfo {author} {\bibfnamefont {D.~A.}\ \bibnamefont {{Smith}}},\
  }\href@noop {} {\bibfield  {journal} {\bibinfo  {journal} {Astrophys. J. L.}\
  }\textbf {\bibinfo {volume} {827}},\ p.~\bibinfo {pages} {39} (\bibinfo
  {year} {2016})}\BibitemShut {NoStop}%
\bibitem [{\citenamefont {{Lasky}}\ \emph {et~al.}(2017)\citenamefont
  {{Lasky}}, \citenamefont {{Leris}}, \citenamefont {{Rowlinson}},\ and\
  \citenamefont {{Glampedakis}}}]{Lasky2017}%
  \BibitemOpen
  \bibfield  {author} {\bibinfo {author} {\bibfnamefont {P.~D.}\ \bibnamefont
  {{Lasky}}}, \bibinfo {author} {\bibfnamefont {C.}~\bibnamefont {{Leris}}},
  \bibinfo {author} {\bibfnamefont {A.}~\bibnamefont {{Rowlinson}}}, \ and\
  \bibinfo {author} {\bibfnamefont {K.}~\bibnamefont {{Glampedakis}}},\
  }\href@noop {} {\bibfield  {journal} {\bibinfo  {journal} {Astrophys. J. L.}\
  }\textbf {\bibinfo {volume} {843}},\ p.~\bibinfo {pages} {1} (\bibinfo {year}
  {2017})}\BibitemShut {NoStop}%
\bibitem [{\citenamefont {{Sarin}}, \citenamefont {{Lasky}},\ and\
  \citenamefont {{Ashton}}(2019{\natexlab{a}})}]{Sarin2019}%
  \BibitemOpen
  \bibfield  {author} {\bibinfo {author} {\bibfnamefont {N.}~\bibnamefont
  {{Sarin}}}, \bibinfo {author} {\bibfnamefont {P.~D.}\ \bibnamefont
  {{Lasky}}}, \ and\ \bibinfo {author} {\bibfnamefont {G.}~\bibnamefont
  {{Ashton}}},\ }\href {\doibase 10.3847/1538-4357/aaf9a0} {\bibfield
  {journal} {\bibinfo  {journal} {Astrophys. J.}\ }\textbf {\bibinfo {volume}
  {872}},\ p.\ \bibinfo {pages} {114} (\bibinfo {year}
  {2019}{\natexlab{a}})}\BibitemShut {NoStop}%
\bibitem [{\citenamefont {Xue}\ \emph {et~al.}(2019)\citenamefont {Xue} \emph
  {et~al.}}]{Xue2019}%
  \BibitemOpen
  \bibfield  {author} {\bibinfo {author} {\bibfnamefont {Y.~Q.}\ \bibnamefont
  {Xue}} \emph {et~al.},\ }\href {https://doi.org/10.1038/s41586-019-1079-5}
  {\bibfield  {journal} {\bibinfo  {journal} {Nature}\ }\textbf {\bibinfo
  {volume} {568}},\ \unskip\ \bibinfo {pages} {198--201} (\bibinfo {year}
  {2019})}\BibitemShut {NoStop}%
\bibitem [{\citenamefont {{Xiao}}, \citenamefont {{Zhang}},\ and\ \citenamefont
  {{Dai}}(2019)}]{Xiao2019}%
  \BibitemOpen
  \bibfield  {author} {\bibinfo {author} {\bibfnamefont {D.}~\bibnamefont
  {{Xiao}}}, \bibinfo {author} {\bibfnamefont {B.-B.}\ \bibnamefont {{Zhang}}},
  \ and\ \bibinfo {author} {\bibfnamefont {Z.-G.}\ \bibnamefont {{Dai}}},\
  }\href@noop {} {\bibfield  {journal} {\bibinfo  {journal} {arXiv e-prints}\
  April} (\bibinfo {year} {2019})},\ \Eprint {http://arxiv.org/abs/1904.05480}
  {arXiv:1904.05480 [astro-ph.HE]} \BibitemShut {NoStop}%
\bibitem [{\citenamefont {{Stratta}}\ \emph {et~al.}(2018)\citenamefont
  {{Stratta}}, \citenamefont {{Dainotti}}, \citenamefont {{Dall'Osso}},
  \citenamefont {{Hernandez}},\ and\ \citenamefont {{De
  Cesare}}}]{Stratta2018}%
  \BibitemOpen
  \bibfield  {author} {\bibinfo {author} {\bibfnamefont {G.}~\bibnamefont
  {{Stratta}}}, \bibinfo {author} {\bibfnamefont {M.~G.}\ \bibnamefont
  {{Dainotti}}}, \bibinfo {author} {\bibfnamefont {S.}~\bibnamefont
  {{Dall'Osso}}}, \bibinfo {author} {\bibfnamefont {X.}~\bibnamefont
  {{Hernandez}}}, \ and\ \bibinfo {author} {\bibfnamefont {G.}~\bibnamefont
  {{De Cesare}}},\ }\href {\doibase 10.3847/1538-4357/aadd8f} {\bibfield
  {journal} {\bibinfo  {journal} {Astrophys. J.}\ }\textbf {\bibinfo {volume}
  {869}},\ p.\ \bibinfo {pages} {155} (\bibinfo {year} {2018})}\BibitemShut
  {NoStop}%
\bibitem [{\citenamefont {{L{\"u}}}, \citenamefont {{Lan}},\ and\ \citenamefont
  {{Liang}}(2019)}]{Lu2019a}%
  \BibitemOpen
  \bibfield  {author} {\bibinfo {author} {\bibfnamefont {H.-J.}\ \bibnamefont
  {{L{\"u}}}}, \bibinfo {author} {\bibfnamefont {L.}~\bibnamefont {{Lan}}}, \
  and\ \bibinfo {author} {\bibfnamefont {E.-W.}\ \bibnamefont {{Liang}}},\
  }\href {\doibase 10.3847/1538-4357/aaf71d} {\bibfield  {journal} {\bibinfo
  {journal} {Astrophy. J.}\ }\textbf {\bibinfo {volume} {871}},\ p.~\bibinfo
  {pages} {54}January (\bibinfo {year} {2019})}\BibitemShut {NoStop}%
\bibitem [{\citenamefont {{Sasmaz Mus}}\ \emph {et~al.}(2019)\citenamefont
  {{Sasmaz Mus}}, \citenamefont {{{\c C}{\i}k{\i}nto{\u g}lu}}, \citenamefont
  {{Aygun}}, \citenamefont {{Ceyhun Anda{\c c}}},\ and\ \citenamefont {{Ek{\c
  s}i}}}]{Mus2019}%
  \BibitemOpen
  \bibfield  {author} {\bibinfo {author} {\bibfnamefont {S.}~\bibnamefont
  {{Sasmaz Mus}}}, \bibinfo {author} {\bibfnamefont {S.}~\bibnamefont {{{\c
  C}{\i}k{\i}nto{\u g}lu}}}, \bibinfo {author} {\bibfnamefont {U.}~\bibnamefont
  {{Aygun}}}, \bibinfo {author} {\bibfnamefont {I.}~\bibnamefont {{Ceyhun
  Anda{\c c}}}}, \ and\ \bibinfo {author} {\bibfnamefont {K.~Y.}\ \bibnamefont
  {{Ek{\c s}i}}},\ }\href@noop {} {\bibfield  {journal} {\bibinfo  {journal}
  {arXiv e-prints}\ } (\bibinfo {year} {2019})},\ \Eprint
  {http://arxiv.org/abs/1904.06769} {arXiv:1904.06769} \BibitemShut {NoStop}%
\bibitem [{\citenamefont {{Burrows}}\ \emph {et~al.}(2005)\citenamefont
  {{Burrows}} \emph {et~al.}}]{Burrows2005}%
  \BibitemOpen
  \bibfield  {author} {\bibinfo {author} {\bibfnamefont {D.~N.}\ \bibnamefont
  {{Burrows}}} \emph {et~al.},\ }\href@noop {} {\bibfield  {journal} {\bibinfo
  {journal} {SSRv}\ }\textbf {\bibinfo {volume} {120}},\ p.\ \bibinfo {pages}
  {165} (\bibinfo {year} {2005})}\BibitemShut {NoStop}%
\bibitem [{\citenamefont {{Sarin}}, \citenamefont {{Lasky}},\ and\
  \citenamefont {{Ashton}}(2019{\natexlab{b}})}]{Sarin2019b}%
  \BibitemOpen
  \bibfield  {author} {\bibinfo {author} {\bibfnamefont {N.}~\bibnamefont
  {{Sarin}}}, \bibinfo {author} {\bibfnamefont {P.~D.}\ \bibnamefont
  {{Lasky}}}, \ and\ \bibinfo {author} {\bibfnamefont {G.}~\bibnamefont
  {{Ashton}}},\ }\href@noop {} {} (\bibinfo {year} {2019}{\natexlab{b}}),\
  \bibinfo {note} {in prep.}\BibitemShut {Stop}%
\bibitem [{\citenamefont {{Berger}}, \citenamefont {{Fong}},\ and\
  \citenamefont {{Chornock}}(2013)}]{Berger2013}%
  \BibitemOpen
  \bibfield  {author} {\bibinfo {author} {\bibfnamefont {E.}~\bibnamefont
  {{Berger}}}, \bibinfo {author} {\bibfnamefont {W.}~\bibnamefont {{Fong}}}, \
  and\ \bibinfo {author} {\bibfnamefont {R.}~\bibnamefont {{Chornock}}},\
  }\href@noop {} {\bibfield  {journal} {\bibinfo  {journal} {Astrophys. J. L.}\
  }\textbf {\bibinfo {volume} {774}},\ p.~\bibinfo {pages} {23} (\bibinfo
  {year} {2013})}\BibitemShut {NoStop}%
\bibitem [{\citenamefont {{Tanvir}}\ \emph {et~al.}(2013)\citenamefont
  {{Tanvir}}, \citenamefont {{Levan}}, \citenamefont {{Fruchter}},
  \citenamefont {{Hjorth}}, \citenamefont {{Hounsell}}, \citenamefont
  {{Wiersema}},\ and\ \citenamefont {{Tunnicliffe}}}]{Tanvir2013}%
  \BibitemOpen
  \bibfield  {author} {\bibinfo {author} {\bibfnamefont {N.~R.}\ \bibnamefont
  {{Tanvir}}}, \bibinfo {author} {\bibfnamefont {A.~J.}\ \bibnamefont
  {{Levan}}}, \bibinfo {author} {\bibfnamefont {A.~S.}\ \bibnamefont
  {{Fruchter}}}, \bibinfo {author} {\bibfnamefont {J.}~\bibnamefont
  {{Hjorth}}}, \bibinfo {author} {\bibfnamefont {R.~A.}\ \bibnamefont
  {{Hounsell}}}, \bibinfo {author} {\bibfnamefont {K.}~\bibnamefont
  {{Wiersema}}}, \ and\ \bibinfo {author} {\bibfnamefont {R.~L.}\ \bibnamefont
  {{Tunnicliffe}}},\ }\href@noop {} {\bibfield  {journal} {\bibinfo  {journal}
  {Nature}\ }\textbf {\bibinfo {volume} {500}},\ p.\ \bibinfo {pages} {547}
  (\bibinfo {year} {2013})}\BibitemShut {NoStop}%
\bibitem [{\citenamefont {{Fan}}, \citenamefont {{Wu}},\ and\ \citenamefont
  {{Wei}}(2013)}]{Fan2013}%
  \BibitemOpen
  \bibfield  {author} {\bibinfo {author} {\bibfnamefont {Y.-Z.}\ \bibnamefont
  {{Fan}}}, \bibinfo {author} {\bibfnamefont {X.-F.}\ \bibnamefont {{Wu}}}, \
  and\ \bibinfo {author} {\bibfnamefont {D.-M.}\ \bibnamefont {{Wei}}},\
  }\href@noop {} {\bibfield  {journal} {\bibinfo  {journal} {Phys. Rev. D}\
  }\textbf {\bibinfo {volume} {88}},\ p.\ \bibinfo {pages} {067304} (\bibinfo
  {year} {2013})}\BibitemShut {NoStop}%
\bibitem [{\citenamefont {{de Ugarte Postigo}}\ \emph
  {et~al.}(2014)\citenamefont {{de Ugarte Postigo}} \emph
  {et~al.}}]{deUgartePostigo2014}%
  \BibitemOpen
  \bibfield  {author} {\bibinfo {author} {\bibfnamefont {A.}~\bibnamefont {{de
  Ugarte Postigo}}} \emph {et~al.},\ }\href@noop {} {\bibfield  {journal}
  {\bibinfo  {journal} {A\&A}\ }\textbf {\bibinfo {volume} {563}},\ p.\
  \bibinfo {pages} {A62} (\bibinfo {year} {2014})}\BibitemShut {NoStop}%
\bibitem [{\citenamefont {{Fong}}\ \emph {et~al.}(2014)\citenamefont {{Fong}}
  \emph {et~al.}}]{Fong2014}%
  \BibitemOpen
  \bibfield  {author} {\bibinfo {author} {\bibfnamefont {W.}~\bibnamefont
  {{Fong}}} \emph {et~al.},\ }\href@noop {} {\bibfield  {journal} {\bibinfo
  {journal} {Astrophys. J.}\ }\textbf {\bibinfo {volume} {780}},\ p.\ \bibinfo
  {pages} {118} (\bibinfo {year} {2014})}\BibitemShut {NoStop}%
\bibitem [{\citenamefont {{Troja}}\ \emph {et~al.}(2016)\citenamefont {{Troja}}
  \emph {et~al.}}]{Troja2016}%
  \BibitemOpen
  \bibfield  {author} {\bibinfo {author} {\bibfnamefont {E.}~\bibnamefont
  {{Troja}}} \emph {et~al.},\ }\href@noop {} {\bibfield  {journal} {\bibinfo
  {journal} {Astrophys. J.}\ }\textbf {\bibinfo {volume} {827}},\ p.\ \bibinfo
  {pages} {102} (\bibinfo {year} {2016})}\BibitemShut {NoStop}%
\bibitem [{\citenamefont {Zhang}\ \emph {et~al.}(2017)\citenamefont {Zhang},
  \citenamefont {Jin}, \citenamefont {Wang},\ and\ \citenamefont
  {Wei}}]{Zhang2017}%
  \BibitemOpen
  \bibfield  {author} {\bibinfo {author} {\bibfnamefont {S.}~\bibnamefont
  {Zhang}}, \bibinfo {author} {\bibfnamefont {Z.-P.}\ \bibnamefont {Jin}},
  \bibinfo {author} {\bibfnamefont {Y.-Z.}\ \bibnamefont {Wang}}, \ and\
  \bibinfo {author} {\bibfnamefont {D.-M.}\ \bibnamefont {Wei}},\ }\href@noop
  {} {\bibfield  {journal} {\bibinfo  {journal} {Astrophys. J. L.}\ }\textbf
  {\bibinfo {volume} {835}},\ p.~\bibinfo {pages} {73} (\bibinfo {year}
  {2017})}\BibitemShut {NoStop}%
\bibitem [{\citenamefont {{Sarin}}\ \emph {et~al.}(2018)\citenamefont
  {{Sarin}}, \citenamefont {{Lasky}}, \citenamefont {{Sammut}},\ and\
  \citenamefont {{Ashton}}}]{Sarin2018}%
  \BibitemOpen
  \bibfield  {author} {\bibinfo {author} {\bibfnamefont {N.}~\bibnamefont
  {{Sarin}}}, \bibinfo {author} {\bibfnamefont {P.~D.}\ \bibnamefont
  {{Lasky}}}, \bibinfo {author} {\bibfnamefont {L.}~\bibnamefont {{Sammut}}}, \
  and\ \bibinfo {author} {\bibfnamefont {G.}~\bibnamefont {{Ashton}}},\
  }\href@noop {} {\bibfield  {journal} {\bibinfo  {journal} {Phys. Rev. D}\
  }\textbf {\bibinfo {volume} {98}},\ p.\ \bibinfo {pages} {043011} (\bibinfo
  {year} {2018})}\BibitemShut {NoStop}%
\bibitem [{\citenamefont {{Melatos}}(1997)}]{Melatos1997}%
  \BibitemOpen
  \bibfield  {author} {\bibinfo {author} {\bibfnamefont {A.}~\bibnamefont
  {{Melatos}}},\ }\href@noop {} {\bibfield  {journal} {\bibinfo  {journal}
  {Mon. Not. R. Astron. Soc.}\ }\textbf {\bibinfo {volume} {288}},\ \unskip\
  \bibinfo {pages} {1049--1059} (\bibinfo {year} {1997})}\BibitemShut {NoStop}%
\bibitem [{\citenamefont {{Antoniadis}}\ \emph {et~al.}(2013)\citenamefont
  {{Antoniadis}} \emph {et~al.}}]{Antoniadis2013}%
  \BibitemOpen
  \bibfield  {author} {\bibinfo {author} {\bibfnamefont {J.}~\bibnamefont
  {{Antoniadis}}} \emph {et~al.},\ }\href@noop {} {\bibfield  {journal}
  {\bibinfo  {journal} {Science}\ }\textbf {\bibinfo {volume} {340}},\ p.\
  \bibinfo {pages} {448} (\bibinfo {year} {2013})}\BibitemShut {NoStop}%
\bibitem [{\citenamefont {{Cromartie}}\ \emph {et~al.}(2019)\citenamefont
  {{Cromartie}} \emph {et~al.}}]{Cromartie2019}%
  \BibitemOpen
  \bibfield  {author} {\bibinfo {author} {\bibfnamefont {H.}~\bibnamefont
  {{Cromartie}}} \emph {et~al.},\ }\href@noop {} {\bibfield  {journal}
  {\bibinfo  {journal} {arXiv e-prints}\ } (\bibinfo {year} {2019})},\ \Eprint
  {http://arxiv.org/abs/1904.06759} {arXiv:1904.06759} \BibitemShut {NoStop}%
\bibitem [{\citenamefont {{Cook}}, \citenamefont {{Shapiro}},\ and\
  \citenamefont {{Teukolsky}}(1994)}]{Cook1994}%
  \BibitemOpen
  \bibfield  {author} {\bibinfo {author} {\bibfnamefont {G.~B.}\ \bibnamefont
  {{Cook}}}, \bibinfo {author} {\bibfnamefont {S.~L.}\ \bibnamefont
  {{Shapiro}}}, \ and\ \bibinfo {author} {\bibfnamefont {S.~A.}\ \bibnamefont
  {{Teukolsky}}},\ }\href@noop {} {\bibfield  {journal} {\bibinfo  {journal}
  {Astrophys. J.}\ }\textbf {\bibinfo {volume} {424}},\ \unskip\ \bibinfo
  {pages} {823--845} (\bibinfo {year} {1994})}\BibitemShut {NoStop}%
\bibitem [{\citenamefont {{Ravi}}\ and\ \citenamefont
  {{Lasky}}(2014)}]{Ravi2014}%
  \BibitemOpen
  \bibfield  {author} {\bibinfo {author} {\bibfnamefont {V.}~\bibnamefont
  {{Ravi}}}\ and\ \bibinfo {author} {\bibfnamefont {P.~D.}\ \bibnamefont
  {{Lasky}}},\ }\href@noop {} {\bibfield  {journal} {\bibinfo  {journal} {Mon.
  Not. R. Astron. Soc.}\ }\textbf {\bibinfo {volume} {441}},\ \unskip\ \bibinfo
  {pages} {2433--2439} (\bibinfo {year} {2014})}\BibitemShut {NoStop}%
\bibitem [{\citenamefont {{Lasky}}\ \emph {et~al.}(2014)\citenamefont
  {{Lasky}}, \citenamefont {{Haskell}}, \citenamefont {{Ravi}}, \citenamefont
  {{Howell}},\ and\ \citenamefont {{Coward}}}]{Lasky2014}%
  \BibitemOpen
  \bibfield  {author} {\bibinfo {author} {\bibfnamefont {P.~D.}\ \bibnamefont
  {{Lasky}}}, \bibinfo {author} {\bibfnamefont {B.}~\bibnamefont {{Haskell}}},
  \bibinfo {author} {\bibfnamefont {V.}~\bibnamefont {{Ravi}}}, \bibinfo
  {author} {\bibfnamefont {E.~J.}\ \bibnamefont {{Howell}}}, \ and\ \bibinfo
  {author} {\bibfnamefont {D.~M.}\ \bibnamefont {{Coward}}},\ }\href@noop {}
  {\bibfield  {journal} {\bibinfo  {journal} {Phys. Rev. D}\ }\textbf {\bibinfo
  {volume} {89}},\ p.\ \bibinfo {pages} {047302} (\bibinfo {year}
  {2014})}\BibitemShut {NoStop}%
\bibitem [{\citenamefont {{Kiziltan}}\ \emph {et~al.}(2013)\citenamefont
  {{Kiziltan}}, \citenamefont {{Kottas}}, \citenamefont {{De Yoreo}},\ and\
  \citenamefont {{Thorsett}}}]{Kiziltan2013}%
  \BibitemOpen
  \bibfield  {author} {\bibinfo {author} {\bibfnamefont {B.}~\bibnamefont
  {{Kiziltan}}}, \bibinfo {author} {\bibfnamefont {A.}~\bibnamefont
  {{Kottas}}}, \bibinfo {author} {\bibfnamefont {M.}~\bibnamefont {{De
  Yoreo}}}, \ and\ \bibinfo {author} {\bibfnamefont {S.~E.}\ \bibnamefont
  {{Thorsett}}},\ }\href@noop {} {\bibfield  {journal} {\bibinfo  {journal}
  {Astrophys. J.}\ }\textbf {\bibinfo {volume} {778}},\ p.~\bibinfo {pages}
  {66} (\bibinfo {year} {2013})}\BibitemShut {NoStop}%
\bibitem [{\citenamefont {{Keitel}}(2019)}]{Keitel2019}%
  \BibitemOpen
  \bibfield  {author} {\bibinfo {author} {\bibfnamefont {D.}~\bibnamefont
  {{Keitel}}},\ }\href {\doibase 10.1093/mnras/stz358} {\bibfield  {journal}
  {\bibinfo  {journal} {Mon. Not. R. Astron. Soc.}\ }\textbf {\bibinfo {volume}
  {485}},\ \unskip\ \bibinfo {pages} {1665--1674} (\bibinfo {year}
  {2019})}\BibitemShut {NoStop}%
\bibitem [{\citenamefont {{Farrow}}, \citenamefont {{Zhu}},\ and\ \citenamefont
  {{Thrane}}(2019)}]{Farrow2019}%
  \BibitemOpen
  \bibfield  {author} {\bibinfo {author} {\bibfnamefont {N.}~\bibnamefont
  {{Farrow}}}, \bibinfo {author} {\bibfnamefont {X.-J.}\ \bibnamefont {{Zhu}}},
  \ and\ \bibinfo {author} {\bibfnamefont {E.}~\bibnamefont {{Thrane}}},\
  }\href@noop {} {\bibfield  {journal} {\bibinfo  {journal} {Astrophys. J.}\ }
  (\bibinfo {year} {2019})},\ \bibinfo {note} {in press},\ \Eprint
  {http://arxiv.org/abs/1902.03300} {arXiv:1902.03300} \BibitemShut {NoStop}%
\bibitem [{\citenamefont {{Evans}}\ \emph {et~al.}(2017)\citenamefont {{Evans}}
  \emph {et~al.}}]{Evans2017}%
  \BibitemOpen
  \bibfield  {author} {\bibinfo {author} {\bibfnamefont {P.~A.}\ \bibnamefont
  {{Evans}}} \emph {et~al.},\ }\href@noop {} {\bibfield  {journal} {\bibinfo
  {journal} {Science}\ }\textbf {\bibinfo {volume} {358}},\ p.\ \bibinfo
  {pages} {1565} (\bibinfo {year} {2017})}\BibitemShut {NoStop}%
\bibitem [{\citenamefont {{Lasky}}\ and\ \citenamefont
  {{Glampedakis}}(2016)}]{Lasky2016}%
  \BibitemOpen
  \bibfield  {author} {\bibinfo {author} {\bibfnamefont {P.~D.}\ \bibnamefont
  {{Lasky}}}\ and\ \bibinfo {author} {\bibfnamefont {K.}~\bibnamefont
  {{Glampedakis}}},\ }\href@noop {} {\bibfield  {journal} {\bibinfo  {journal}
  {Mon. Not. R. Astron. Soc.}\ }\textbf {\bibinfo {volume} {458}},\ \unskip\
  \bibinfo {pages} {1660--1670} (\bibinfo {year} {2016})}\BibitemShut {NoStop}%
\bibitem [{\citenamefont {{Rowlinson}}\ \emph {et~al.}(2014)\citenamefont
  {{Rowlinson}}, \citenamefont {{Gompertz}}, \citenamefont {{Dainotti}},
  \citenamefont {{O'Brien}}, \citenamefont {{Wijers}},\ and\ \citenamefont
  {{van der Horst}}}]{Rowlinson2014}%
  \BibitemOpen
  \bibfield  {author} {\bibinfo {author} {\bibfnamefont {A.}~\bibnamefont
  {{Rowlinson}}}, \bibinfo {author} {\bibfnamefont {B.~P.}\ \bibnamefont
  {{Gompertz}}}, \bibinfo {author} {\bibfnamefont {M.}~\bibnamefont
  {{Dainotti}}}, \bibinfo {author} {\bibfnamefont {P.~T.}\ \bibnamefont
  {{O'Brien}}}, \bibinfo {author} {\bibfnamefont {R.~A.~M.~J.}\ \bibnamefont
  {{Wijers}}}, \ and\ \bibinfo {author} {\bibfnamefont {A.~J.}\ \bibnamefont
  {{van der Horst}}},\ }\href@noop {} {\bibfield  {journal} {\bibinfo
  {journal} {Mon. Not. R. Astron. Soc.}\ }\textbf {\bibinfo {volume} {443}},\
  \unskip\ \bibinfo {pages} {1779--1787} (\bibinfo {year} {2014})}\BibitemShut
  {NoStop}%
\bibitem [{\citenamefont {{LIGO Scientific Collaboration}}(2007)}]{PSD:iLIGO}%
  \BibitemOpen
  \bibfield  {author} {\bibinfo {author} {\bibnamefont {{LIGO Scientific
  Collaboration}}},\ }\href@noop {} {\bibinfo {title} {{Advanced LIGO
  anticipated sensitivity curves}},\ }\  \bibinfo {year} {2007} \unskip,\
  \bibinfo {note} {https://dcc.ligo.org/LIGO-T1100338/public}\BibitemShut
  {NoStop}%
\bibitem [{\citenamefont {{LIGO Scientific Collaboration}}(2018)}]{PSD:ALIGO}%
  \BibitemOpen
  \bibfield  {author} {\bibinfo {author} {\bibnamefont {{LIGO Scientific
  Collaboration}}},\ }\href@noop {} {\bibinfo {title} {{Updated Advanced LIGO
  sensitivity design curve}},\ }\  \bibinfo {year} {2018} \unskip,\ \bibinfo
  {note} {https://dcc.ligo.org/T1800044-v5}\BibitemShut {NoStop}%
\bibitem [{\citenamefont {{Punturo}}\ \emph {et~al.}(2010)\citenamefont
  {{Punturo}} \emph {et~al.}}]{Punturo2010}%
  \BibitemOpen
  \bibfield  {author} {\bibinfo {author} {\bibfnamefont {M.}~\bibnamefont
  {{Punturo}}} \emph {et~al.},\ }\href@noop {} {\bibfield  {journal} {\bibinfo
  {journal} {Class. and Q. Grav.}\ }\textbf {\bibinfo {volume} {27}},\ p.\
  \bibinfo {pages} {084007} (\bibinfo {year} {2010})}\BibitemShut {NoStop}%
\bibitem [{\citenamefont {{Hild}}\ \emph {et~al.}(2011)\citenamefont {{Hild}}
  \emph {et~al.}}]{Hild2011}%
  \BibitemOpen
  \bibfield  {author} {\bibinfo {author} {\bibfnamefont {S.}~\bibnamefont
  {{Hild}}} \emph {et~al.},\ }\href@noop {} {\bibfield  {journal} {\bibinfo
  {journal} {Classical and Quantum Gravity}\ }\textbf {\bibinfo {volume}
  {28}},\ p.\ \bibinfo {pages} {094013} (\bibinfo {year} {2011})}\BibitemShut
  {NoStop}%
\bibitem [{\citenamefont {Bonazzola}\ and\ \citenamefont
  {Gourgoulhon}(1996)}]{Bonazzola1996}%
  \BibitemOpen
  \bibfield  {author} {\bibinfo {author} {\bibfnamefont {S.}~\bibnamefont
  {Bonazzola}}\ and\ \bibinfo {author} {\bibfnamefont {E.}~\bibnamefont
  {Gourgoulhon}},\ }\href@noop {} {\bibfield  {journal} {\bibinfo  {journal}
  {A\&A}\ }\textbf {\bibinfo {volume} {312}},\ p.\ \bibinfo {pages} {675}
  (\bibinfo {year} {1996})}\BibitemShut {NoStop}%
\bibitem [{\citenamefont {Cutler}(2002)}]{Cutler2002}%
  \BibitemOpen
  \bibfield  {author} {\bibinfo {author} {\bibfnamefont {C.}~\bibnamefont
  {Cutler}},\ }\href@noop {} {\bibfield  {journal} {\bibinfo  {journal} {Phys.
  Rev. D}\ }\textbf {\bibinfo {volume} {66}},\ p.\ \bibinfo {pages} {084025}
  (\bibinfo {year} {2002})}\BibitemShut {NoStop}%
\bibitem [{\citenamefont {{Lai}}\ and\ \citenamefont
  {{Shapiro}}(1995)}]{Lai1995}%
  \BibitemOpen
  \bibfield  {author} {\bibinfo {author} {\bibfnamefont {D.}~\bibnamefont
  {{Lai}}}\ and\ \bibinfo {author} {\bibfnamefont {S.~L.}\ \bibnamefont
  {{Shapiro}}},\ }\href@noop {} {\bibfield  {journal} {\bibinfo  {journal}
  {Astrophys. J.}\ }\textbf {\bibinfo {volume} {442}},\ \unskip\ \bibinfo
  {pages} {259--272} (\bibinfo {year} {1995})}\BibitemShut {NoStop}%
\bibitem [{\citenamefont {{Andersson}}(1998)}]{Andersson1998}%
  \BibitemOpen
  \bibfield  {author} {\bibinfo {author} {\bibfnamefont {N.}~\bibnamefont
  {{Andersson}}},\ }\href@noop {} {\bibfield  {journal} {\bibinfo  {journal}
  {Astrophys. J.}\ }\textbf {\bibinfo {volume} {502}},\ \unskip\ \bibinfo
  {pages} {708--713} (\bibinfo {year} {1998})}\BibitemShut {NoStop}%
\bibitem [{\citenamefont {{Corsi}}\ and\ \citenamefont
  {{M{\'e}sz{\'a}ros}}(2009)}]{Corsi2009}%
  \BibitemOpen
  \bibfield  {author} {\bibinfo {author} {\bibfnamefont {A.}~\bibnamefont
  {{Corsi}}}\ and\ \bibinfo {author} {\bibfnamefont {P.}~\bibnamefont
  {{M{\'e}sz{\'a}ros}}},\ }\href@noop {} {\bibfield  {journal} {\bibinfo
  {journal} {Astrophys. J.}\ }\textbf {\bibinfo {volume} {702}},\ \unskip\
  \bibinfo {pages} {1171--1178} (\bibinfo {year} {2009})}\BibitemShut {NoStop}%
\bibitem [{\citenamefont {{Dall'Osso}}\ \emph {et~al.}(2015)\citenamefont
  {{Dall'Osso}}, \citenamefont {{Giacomazzo}}, \citenamefont {{Perna}},\ and\
  \citenamefont {{Stella}}}]{DallOsso2015}%
  \BibitemOpen
  \bibfield  {author} {\bibinfo {author} {\bibfnamefont {S.}~\bibnamefont
  {{Dall'Osso}}}, \bibinfo {author} {\bibfnamefont {B.}~\bibnamefont
  {{Giacomazzo}}}, \bibinfo {author} {\bibfnamefont {R.}~\bibnamefont
  {{Perna}}}, \ and\ \bibinfo {author} {\bibfnamefont {L.}~\bibnamefont
  {{Stella}}},\ }\href@noop {} {\bibfield  {journal} {\bibinfo  {journal}
  {Astrophys. J.}\ }\textbf {\bibinfo {volume} {798}},\ p.~\bibinfo {pages}
  {25} (\bibinfo {year} {2015})}\BibitemShut {NoStop}%
\bibitem [{\citenamefont {{Doneva}}, \citenamefont {{Kokkotas}},\ and\
  \citenamefont {{Pnigouras}}(2015)}]{Doneva2015}%
  \BibitemOpen
  \bibfield  {author} {\bibinfo {author} {\bibfnamefont {D.~D.}\ \bibnamefont
  {{Doneva}}}, \bibinfo {author} {\bibfnamefont {K.~D.}\ \bibnamefont
  {{Kokkotas}}}, \ and\ \bibinfo {author} {\bibfnamefont {P.}~\bibnamefont
  {{Pnigouras}}},\ }\href@noop {} {\bibfield  {journal} {\bibinfo  {journal}
  {Phys. Rev. D}\ }\textbf {\bibinfo {volume} {92}},\ p.\ \bibinfo {pages}
  {104040} (\bibinfo {year} {2015})}\BibitemShut {NoStop}%
\bibitem [{\citenamefont {{Gao}}, \citenamefont {{Cao}},\ and\ \citenamefont
  {{Zhang}}(2017)}]{Gao2017}%
  \BibitemOpen
  \bibfield  {author} {\bibinfo {author} {\bibfnamefont {H.}~\bibnamefont
  {{Gao}}}, \bibinfo {author} {\bibfnamefont {Z.}~\bibnamefont {{Cao}}}, \ and\
  \bibinfo {author} {\bibfnamefont {B.}~\bibnamefont {{Zhang}}},\ }\href@noop
  {} {\bibfield  {journal} {\bibinfo  {journal} {Astrophys. J.}\ }\textbf
  {\bibinfo {volume} {844}},\ p.\ \bibinfo {pages} {112} (\bibinfo {year}
  {2017})}\BibitemShut {NoStop}%
\bibitem [{\citenamefont {{Dai}}(2019)}]{Dai2019}%
  \BibitemOpen
  \bibfield  {author} {\bibinfo {author} {\bibfnamefont {Z.~G.}\ \bibnamefont
  {{Dai}}},\ }\href@noop {} {\bibfield  {journal} {\bibinfo  {journal} {A\&A}\
  }\textbf {\bibinfo {volume} {622}},\ p.\ \bibinfo {pages} {A194} (\bibinfo
  {year} {2019})}\BibitemShut {NoStop}%
\bibitem [{\citenamefont {{Piro}}, \citenamefont {{Giacomazzo}},\ and\
  \citenamefont {{Perna}}(2017)}]{Piro2017}%
  \BibitemOpen
  \bibfield  {author} {\bibinfo {author} {\bibfnamefont {A.~L.}\ \bibnamefont
  {{Piro}}}, \bibinfo {author} {\bibfnamefont {B.}~\bibnamefont
  {{Giacomazzo}}}, \ and\ \bibinfo {author} {\bibfnamefont {R.}~\bibnamefont
  {{Perna}}},\ }\href@noop {} {\bibfield  {journal} {\bibinfo  {journal}
  {Astrophys. J. L.}\ }\textbf {\bibinfo {volume} {844}},\ p.\ \bibinfo {pages}
  {L19} (\bibinfo {year} {2017})}\BibitemShut {NoStop}%
\bibitem [{\citenamefont {{Li}}\ \emph {et~al.}(2016)\citenamefont {{Li}},
  \citenamefont {{Zhang}}, \citenamefont {{Zhang}}, \citenamefont {{Gao}},
  \citenamefont {{Qi}},\ and\ \citenamefont {{Liu}}}]{Li2016}%
  \BibitemOpen
  \bibfield  {author} {\bibinfo {author} {\bibfnamefont {A.}~\bibnamefont
  {{Li}}}, \bibinfo {author} {\bibfnamefont {B.}~\bibnamefont {{Zhang}}},
  \bibinfo {author} {\bibfnamefont {N.-B.}\ \bibnamefont {{Zhang}}}, \bibinfo
  {author} {\bibfnamefont {H.}~\bibnamefont {{Gao}}}, \bibinfo {author}
  {\bibfnamefont {B.}~\bibnamefont {{Qi}}}, \ and\ \bibinfo {author}
  {\bibfnamefont {T.}~\bibnamefont {{Liu}}},\ }\href@noop {} {\bibfield
  {journal} {\bibinfo  {journal} {Phys. Rev. D}\ }\textbf {\bibinfo {volume}
  {94}},\ p.\ \bibinfo {pages} {083010} (\bibinfo {year} {2016})}\BibitemShut
  {NoStop}%
\bibitem [{\citenamefont {{Li}}, \citenamefont {{Zhu}},\ and\ \citenamefont
  {{Zhou}}(2017)}]{Li2017}%
  \BibitemOpen
  \bibfield  {author} {\bibinfo {author} {\bibfnamefont {A.}~\bibnamefont
  {{Li}}}, \bibinfo {author} {\bibfnamefont {Z.-Y.}\ \bibnamefont {{Zhu}}}, \
  and\ \bibinfo {author} {\bibfnamefont {X.}~\bibnamefont {{Zhou}}},\
  }\href@noop {} {\bibfield  {journal} {\bibinfo  {journal} {Astrophys. J.}\
  }\textbf {\bibinfo {volume} {844}},\ p.~\bibinfo {pages} {41} (\bibinfo
  {year} {2017})}\BibitemShut {NoStop}%
\bibitem [{\citenamefont {{Drago}}\ and\ \citenamefont
  {{Pagliara}}(2018)}]{Drago2018}%
  \BibitemOpen
  \bibfield  {author} {\bibinfo {author} {\bibfnamefont {A.}~\bibnamefont
  {{Drago}}}\ and\ \bibinfo {author} {\bibfnamefont {G.}~\bibnamefont
  {{Pagliara}}},\ }\href {\doibase 10.3847/2041-8213/aaa40a} {\bibfield
  {journal} {\bibinfo  {journal} {Astrophys. J. L.}\ }\textbf {\bibinfo
  {volume} {852}},\ p.\ \bibinfo {pages} {L32} (\bibinfo {year}
  {2018})}\BibitemShut {NoStop}%
\bibitem [{\citenamefont {{Gao}}, \citenamefont {{Zhang}},\ and\ \citenamefont
  {{L{\"u}}}(2016)}]{Gao2016}%
  \BibitemOpen
  \bibfield  {author} {\bibinfo {author} {\bibfnamefont {H.}~\bibnamefont
  {{Gao}}}, \bibinfo {author} {\bibfnamefont {B.}~\bibnamefont {{Zhang}}}, \
  and\ \bibinfo {author} {\bibfnamefont {H.-J.}\ \bibnamefont {{L{\"u}}}},\
  }\href@noop {} {\bibfield  {journal} {\bibinfo  {journal} {Phys. Rev. D}\
  }\textbf {\bibinfo {volume} {93}},\ p.\ \bibinfo {pages} {044065} (\bibinfo
  {year} {2016})}\BibitemShut {NoStop}%
\bibitem [{\citenamefont {{L{\"u}}}\ \emph {et~al.}(2017)\citenamefont
  {{L{\"u}}}, \citenamefont {{Zhang}}, \citenamefont {{Zhong}}, \citenamefont
  {{Hou}}, \citenamefont {{Sun}}, \citenamefont {{Rice}},\ and\ \citenamefont
  {{Liang}}}]{Lu2017}%
  \BibitemOpen
  \bibfield  {author} {\bibinfo {author} {\bibfnamefont {H.-J.}\ \bibnamefont
  {{L{\"u}}}}, \bibinfo {author} {\bibfnamefont {H.-M.}\ \bibnamefont
  {{Zhang}}}, \bibinfo {author} {\bibfnamefont {S.-Q.}\ \bibnamefont
  {{Zhong}}}, \bibinfo {author} {\bibfnamefont {S.-J.}\ \bibnamefont {{Hou}}},
  \bibinfo {author} {\bibfnamefont {H.}~\bibnamefont {{Sun}}}, \bibinfo
  {author} {\bibfnamefont {J.}~\bibnamefont {{Rice}}}, \ and\ \bibinfo {author}
  {\bibfnamefont {E.-W.}\ \bibnamefont {{Liang}}},\ }\href@noop {} {\bibfield
  {journal} {\bibinfo  {journal} {Astrophys. J.}\ }\textbf {\bibinfo {volume}
  {835}},\ p.\ \bibinfo {pages} {181} (\bibinfo {year} {2017})}\BibitemShut
  {NoStop}%
\bibitem [{\citenamefont {{Lin}}\ and\ \citenamefont {{Lu}}(2019)}]{Lin2019}%
  \BibitemOpen
  \bibfield  {author} {\bibinfo {author} {\bibfnamefont {J.}~\bibnamefont
  {{Lin}}}\ and\ \bibinfo {author} {\bibfnamefont {R.-J.}\ \bibnamefont
  {{Lu}}},\ }\href@noop {} {\bibfield  {journal} {\bibinfo  {journal}
  {Astrophys. J.}\ }\textbf {\bibinfo {volume} {871}},\ p.\ \bibinfo {pages}
  {160} (\bibinfo {year} {2019})}\BibitemShut {NoStop}%
\bibitem [{\citenamefont {{Lyford}}, \citenamefont {{Baumgarte}},\ and\
  \citenamefont {{Shapiro}}(2003)}]{Lyford2003}%
  \BibitemOpen
  \bibfield  {author} {\bibinfo {author} {\bibfnamefont {N.~D.}\ \bibnamefont
  {{Lyford}}}, \bibinfo {author} {\bibfnamefont {T.~W.}\ \bibnamefont
  {{Baumgarte}}}, \ and\ \bibinfo {author} {\bibfnamefont {S.~L.}\ \bibnamefont
  {{Shapiro}}},\ }\href@noop {} {\bibfield  {journal} {\bibinfo  {journal}
  {Astrophys. J.}\ }\textbf {\bibinfo {volume} {583}},\ \unskip\ \bibinfo
  {pages} {410--415} (\bibinfo {year} {2003})}\BibitemShut {NoStop}%
\end{thebibliography}%

\end{document}